\documentclass[12pt,american]{paper}

\usepackage[utf8]{inputenc}
\usepackage{geometry}
\usepackage{amsmath}
\usepackage{amssymb}
\usepackage{graphicx}
\usepackage{esint}
\usepackage[authoryear]{natbib}

\makeatletter
\usepackage{bbm}\usepackage[font=small,format=plain,labelfont=bf,up,textfont=up]{caption}

\usepackage{graphicx}

\renewcommand{\neg}{\mathord{\sim}}

\usepackage{amsthm}\theoremstyle{definition}

\makeatletter
\renewcommand*\l@section{\@dottedtocline{1}{0em}{1.5em}}
\renewcommand*\l@subsection{\@dottedtocline{2}{1.5em}{2.3em}}
\renewcommand*\l@subsubsection{\@dottedtocline{3}{4.3em}{3.2em}}
\makeatother

\makeatother

\usepackage{babel}
\begin{document}

\title{Quantum mechanics as a deterministic theory of a continuum of worlds}

\author{Kim Joris Boström}
\maketitle
\begin{abstract}
A non-relativistic quantum mechanical theory is proposed that describes
the universe as a continuum of worlds whose mutual interference gives
rise to quantum phenomena. A logical framework is introduced to properly
deal with propositions about objects in a multiplicity of worlds.
In this logical framework, the continuum of worlds is treated in analogy
to the continuum of time points; both ``time'' and ``world'' are
considered as mutually independent modes of existence. The theory
combines elements of Bohmian mechanics and of Everett's many-worlds
interpretation; it has a clear ontology and a set of precisely defined
postulates from where the predictions of standard quantum mechanics
can be derived. Probability as given by the Born rule emerges as a
consequence of insufficient knowledge of observers about which world
it is that they live in. The theory describes a continuum of worlds
rather than a single world or a discrete set of worlds, so it is similar
in spirit to many-worlds interpretations based on Everett's approach,
without being actually reducible to these. In particular, there is
no splitting of worlds, which is a typical feature of Everett-type
theories. Altogether, the theory explains (1) the subjective occurrence
of probabilities, (2) their quantitative value as given by the Born
rule, and (3) the apparently random ``collapse of the wavefunction''
caused by the measurement, while still being an objectively deterministic
theory. \end{abstract}
\tableofcontents{}

\section{Introduction\label{sec:Introduction}}

The ideas proposed in this paper have grown out of dissatisfaction
with the existing interpretations of quantum mechanics. The problem
is not that quantum mechanics does not yield the correct experimental
\textit{predictions}, but rather that there is still no consensus
about the \textit{metaphysical content} of the theory, that is, the
story that quantum mechanics tells us about reality. Some people simply
turn the tables and consider the lack of a clear, indisputable metaphysical
interpretation not as a \textit{bug} but rather a \textit{feature},
denying the existence of objective reality altogether. Doing so, however,
appears to me as an act of resignation rather than a satisfying solution
to the conundrum.

\begin{figure}
\includegraphics[width=1\textwidth]{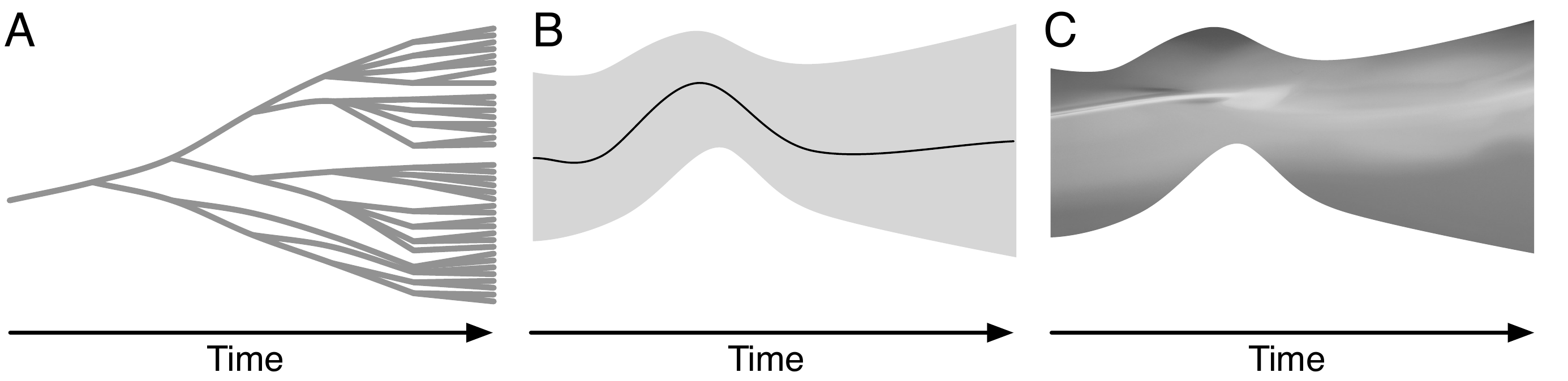}

\caption{\label{fig:Ontologies}Schematic illustration of the different ontologies
of Everettian mechanics (A), Bohmian mechanics (B), and the here-proposed
theory (C). Everettian mechanics describes a universe of constantly
branching, vaguely-defined worlds that evolve through Hilbert space.
Bohmian mechanics describes a wavefunction plus one single precisely-defined
world that both evolve through configuration space. The here-proposed
theory describes the a continuum of precisely-defined worlds of varying
density, which evolves through configuration space, and where each
world corresponds to a Bohmian trajectory. The wavefunction serves
as a generating function for the continuum of trajectories, which
constitutes the physically existing universe.}

\end{figure}

The theory that I am going to propose offers a transparent and consistent
interpretation of non-relativistic quantum mechanics. Measurements
are taken to be ordinary processes, there is no objective ``collapse
of the wavefunction'', and the wavefunction is a convenient mathematical
representation of a physically existing continuum of trajectories
through configuration space, each one corresponding to an individual
\textit{world}. There is no distinction between a ``quantum system''
and a ``classical apparatus'' to explain the definite outcome of
measurements and their probabilistic behavior. Elementary particles
have at all times and in all worlds a well-defined position in 3D
space; there are no such things as ``probability clouds'' and ``objective
uncertainty''. Rather, probability emerges as the consequence of
insufficient knowledge of observers about which world it is that they
live in. The quantitative form of such epistemic probability does
not rely on a ``quantum equilibrium hypothesis'' as in Bohmian mechanics,
or a ``branch weight'' as in Everettian mechanics, but is derived
from the concept of a \textit{substantial density} of trajectories
in configuration space, which is regarded as an objective feature
of the physically existing universe (or \textit{multiverse}, if one
prefers to say). More precisely, the universe (multiverse) is conceived
of as a continuum of trajectories endowed with a certain density that
determines how densely the trajectories are packed in different regions
of configuration space. Each trajectory corresponds to a \textit{world},
and all worlds equally exist. My proposal is different from Bohmian
mechanics in that the wavefunction does not represent a physical field
existing in addition to particles, and it is different from Everettian
mechanics in that the worlds are precisely defined and do not split
(Figure \ref{fig:Ontologies}). The theory is based on ideas initially
published as a preprint draft \citep{Bostrom2012}, which has been
completely re-worked and enhanced, in particular by adding a logical
framework to properly deal with propositions about physical systems
in a multiplicity of worlds, and by providing the conceptual prerequisites
for treating the collection of worlds as a continuous substance. After
having finished and submitted an earlier version of this manuscript,
I noticed that essentially the same theory, though with a stronger
focus on formal aspects and less focus on ontological and epistemological
matters, has independently been put forward by Poirier and Schiff
\citep{Poirier2010,Schiff_et_al_2012}. Although already having been
aware of, and having cited, these publications, I did not fully recognize
how close their theory was to mine. The similarities have probably
been masked from my eyes by the authors' central emphasis on the elimination
of the wavefunction from the theory, which was (and is) not an issue
in my approach. The wavefunction in my approach is a generating function
of, and thus a mathematical representative for, a continuum of trajectories
identified as worlds, and so it has its own justification to remain
within the theory. I will discuss the approach by Poirier and Schiff
in some more detail in the last section, along with related approaches
by \citet{Tipler2006}, \citet{Hall_et_al_2014}, and \citet{Sebens2014}.

The remaining part of the paper is structured as follows. In section
\ref{sec:Logical-framework}, a logical framework is proposed to properly
deal with propositions evaluated in a multiplicity of worlds and at
a multiplicity of times, whereby \textit{time} and \textit{world}
are treated as so-called \textit{modes of existence}. The concept
of a \textit{world continuum} is introduced as a physically existing
entity consisting of a continuum of worlds. Also, the concept of an
\textit{instance} is introduced, which comes in three flavors. \textit{A
time-instance} is an instance of an object at a specific time, a \textit{world-instance}
is an instance of an object in a specific world, and a \textit{time-world-instance}
is an instance of an object at both a specific time and in a specific
world. The logical framework is then applied to the case of particles
as the basic entities of the theory. It is postulated that each world
corresponds to one and only one trajectory of the universe in configuration
space, so that worlds can be identified with trajectories. Each time-world-instance
of the universe corresponds to the configuration of all particles
in the universe at a specific time and in a specific world. 

In section \ref{sec:Probability}, the notion of \textit{continuous
substance} is introduced and discussed to properly deal with a continuum
of worlds later on. In particular, the \textit{substantial amount
}and the \textit{substantial density} of a given substance are defined
and discussed. Subsequently, probability is put forward as an epistemic
concept deriving from the insufficient knowledge of observers about
which world it is that they live in. To this aim, the well-known Laplacian
rule is generalized to a continuous number of possibilities with the
help of a measure that is specified as the substantial amount of worlds
whose trajectories are crossing a certain region in configuration
space. 

The role and the physical meaning of the wavefunction is discussed
in section \ref{sec:Wavefunction}, as well as the role and the meaning
of configuration space and of the world continuum. It is held that
the wavefunction does not represent a physically existing entity itself
but is rather to be considered an abstract generating function for
the physically existing world continuum. 

A minimal set of postulates is given in section \ref{sec:Foundations}
to formally define the theory. Measurement is introduced as a special
case of an otherwise ordinary interaction between a system of interest
and a measurement apparatus, yielding the Born rule as a subjective
measure of uncertainty about measurement outcomes obtained in individual
worlds. The ``collapse of the wavefunction'' is derived as a useful
but merely subjective description at the level of an individual world. 

In the final section, several aspects of my proposal and their relation
to other proposals and criticisms found in the literature are discussed.
In particular, the relation between objective reality and subjective
experience in the presence of a multiplicity of worlds is addressed.
My proposal is compared to Bohmian mechanics, to Tipler's formulation
of quantum mechanics \citep{Tipler2006}, to the MIW approach of Hall
et al. \citep{Hall_et_al_2014}, to Sebens' Newtonian QM \citep{Sebens2014},
and to Poirier and Schiff's approach \citep{Poirier2010,Schiff_et_al_2012}.
I will also respond to criticisms raised by \citet{Vaidman2014} and
\citet{Sebens2014} against the idea of a continuum of worlds.

\section{Logical framework\label{sec:Logical-framework}}

After having long been ignored and ridiculed, the many-worlds interpretation
has in recent years become a scientifically recognized and intensely
discussed interpretation of quantum mechanics \citep[see][for modern accounts]{Tegmark1998,Vaidman2008,Wallace2008}.
Among the main objections against the many-worlds interpretation is
the criticism that it is too vague about the notion of \textit{worlds},
and that the entire conception of many worlds existing in parallel
is absurd in the first place \citep[cf.][]{Kent1990,Barrett2011}.
Indeed, at first sight it appears problematic or even absurd to consider
our world as one out of many worlds, for we usually speak of \emph{the}
world and understand it as \textit{everything that exists}. The inventor
of many-worlds quantum theory, Hugh Everett III, originally did not
use the term ``world'', and he named his theory the ``relative-state
formulation of quantum mechanics''~\citep{Everett1957} and ``the
theory of the universal wavefunction'' \citep{Everett1973}. However,
the notion of \textit{worlds} became more and more popular in discussions
of Everett's theory \citep[cf.][]{Werner1962,Barrett2011}, and was
eventually introduced to the public by DeWitt~\citep{DeWitt1970}.
I will stick to the now firmly established convention of using the
term ``world'', though I will use it in a very specific manner.
In the extended logical framework that I will propose, there are two
\textit{modes of existence}, and these are \textit{time} and \textit{world}.
The key idea is very simple. In conventional logic, objects cannot
\textit{at the same time} have a property and not have that property.
In the proposed logical framework, objects cannot \textit{at the same
time and in the same world} have a property and not have that property.
So, while in conventional logic there is only one mode of existence,
which is time, in the extended framework there are two modes of existence,
which are \textit{time} and \textit{world.} We will first concentrate
on the \textit{time} mode and suppress the \textit{world} mode. Once
we have clarified how to treat \textit{time} in a logical framework,
and once we have set up the terminology, we can straightforwardly
extend the framework to also include the \textit{world }mode. We shall
see that \textit{world} and \textit{time} are treated in an analogous
manner, and since we are already quite familiar with the concept of
\textit{time} – at least we have some intuition in using this concept
– we will be able to understand how the concept of \textit{world}
can be understood as well. In a similar sense that classical mechanics
can be understood as a \textit{many-times} theory, non-relativistic
quantum mechanics can be understood as a \textit{many-worlds }theory
in addition to its many-times aspect.

Why should one take it that quantum mechanics, in contrast to classical
mechanics, needs a \textit{world} mode of existence in addition to
the \textit{time }mode? Because in quantum mechanics the worlds \textit{interfere}
with each other. This interference of worlds is responsible for the
typical ``quantum'' phenomena that go beyond classical explanation.
For example, an electron going through a double-slit produces an interference
pattern because its trajectory in one world interferes with its trajectories
in other worlds. In the classical theory there are no interference
patterns produced by individual particles, so there is no need to
consider additional worlds that interfere with each other. Other experimental
paradigms where a many-worlds interpretation yields a straightforward
and transparent explanation is neutron interference \citep{Vaidman1998},
quantum computation \citep{Deutsch2012} and counterfactual measurement
\citep{Elitzur_et_al_1993,Vaidman1994,Vaidman2009}\textit{.} After
all, the most important reason to favor a many-worlds interpretation
is not simply to satisfy a somewhat romantic attitude but rather to
better understand quantum phenomena and to avoid serious conceptual
difficulties.

\subsection{General formulation}

Let us begin with the more familiar concept of \textit{time}. Physically
existing objects have their properties each \textit{at a given instance
of time}. An instance of time is also referred to as a \textit{time
point}, but in the following we shall drop the explicit distinction
between \textit{time points} and \textit{times} whenever there is
no risk of confusion. That is, by saying that an object \textit{$o$
}has a particular property \textit{$F$} at a given \textit{time}
$t$, we mean that the object has the property at the \textit{time
point} $t$. We denote this statement by the formula $p(t)$=``$Fo\,@t$''
and call it an \textit{anchored proposition.} As usual in a physical
theory, time points are represented by real numbers, so the set $\mathcal{T}$
of all time points equals the entire real line, $\mathcal{T}=\mathbbm R$.
The temporal anchor plays the role of the \textit{present time }of
an anchored proposition\textit{. }Consider an \textit{unanchored}
proposition like ``the traffic lights are green''. In one moment
the traffic lights may be green, in another moment they may be red,
or yellow, or something else. We can anchor an unanchored proposition
by adding ``now'' to it, where ``now'' is an \textit{unspecified
temporal anchor} to be treated as a variable. Formally, the partial
expression ``$Fo$'' occurring in the proposition $p(t)$=``$Fo\,@t$''
is the \textit{body} of the proposition, and $t$ is the \textit{(temporal)
anchor}. If the temporal anchor is unspecified, that is, it is not
given by a real number but remains a variable, then the proposition
remains \textit{unevaluated}. We can evaluate it either by specifying
a numerical value for $t$, or by quantifying over $t$, so that,
for example, for a given time period $T\subset\mathcal{T}$ the proposition
``$\forall(t\in T):\ Fo\,@t$'' is understood as ``object $o$
has property $F$ during $T$'', which is either true or false. The
proposition is then anchored to the specified interval $T$, which
may also coincide with the entire time domain $\mathcal{T}$. In the
latter case we may use the shorthand notation ``$\forall t$'' to
mean ``$\forall(t\in\mathcal{T})$'', and similarly ``$\exists t$''
to mean ``$\exists(t\in\mathcal{T})$''.

Since real numbers form an ordered set, a temporal anchor $t$ can
be set into relation with other time points, so that time points that
are smaller or bigger than $t$, in the usual order of $\mathbbm R$,
play the role of, respectively \textit{past (earlier) }or \textit{future
(later)} time points relative to $t$. We then can then formalize
the proposition ``Object $o$ will have the property $F$'' as ``$\exists(t'>t):~Fo\,@t'$'',
and the proposition ``Object $o$ never had the property $F$''
as ``$\forall(t'<t):~\neg Fo\,@t'$''. The variable $t$ is then
the unspecified anchor of both propositions and represents the ``now''
of the proposition, that is, the present time relative to which the
body of each proposition is anchored in the future and the past, respectively.

Let us now extend the logical framework by adding \textit{world} as
another mode of existence. That is, an object has properties only
with respect to a particular time and a particular world. Anchored
propositions are then of the form $p(t,w)$=``$Fo\,@(t,w)$'', which
is understood as ``object $o$ has property $F$ at time $t$ and
in world $w$''. The expression ``$Fo$'' is, again, the \textit{body
}of the anchored proposition $p(t,w)$, and the expression ``$@(t,w)$''
is the \textit{anchor}, with $t$ being the \textit{temporal anchor}
and $w$ being the \textit{world anchor.} Just like the temporal anchor
$t$ plays the role of the \textit{present time} of the proposition,
the world anchor $w$ plays the role of the \textit{actual world}
of the proposition\textit{. }Let us denote the set of all worlds by
$\mathcal{W}$ and allow to quantify over both worlds and times, so
that, say, for a given set $W\subset\mathcal{W}$ the formal proposition
$p(t)$=``$\forall(w\in W)\exists(t'<t):\ Fo\,@(t',w)$'' reads
``for all worlds in $W$ the object $o$ once had the property $F$''.
That proposition involves the variable $t$ as an unspecified temporal
anchor, and it remains unevaluated until $t$ is specified or quantified.
Consider another different example involving the variable $w$ as
an unspecified world anchor: the formal proposition $p(w)$=``$\forall t:\ Fo\,@(t,w)$''
reads ``in the actual world the object $o$ always has the property
$F$'' and remains unevaluated as long as the actual world $w$ is
not specified or quantified. If the actual world of the proposition
lies within some given set $W\subset\mathcal{W}$, then the proposition
becomes $p$=``$\forall t\forall(w\in W):\ Fo\,@(t,w)$'', which
is understood as ``in all worlds contained within $W$ the object
$o$ always has the property $F$''. The proposition $p$ contains
no free variables and hence is an evaluated proposition. In the case
$W=\mathcal{W}$ we may shorten ``$\forall(w\in\mathcal{W})$''
into ``$\forall w$'', and ``$\exists(w\in\mathcal{W})$'' into
``$\exists w$''. 

There is yet another, logically equivalent, way of understanding anchored
propositions, which might be favorable under certain circumstances.
The occurrence of an object at a particular time and in a particular
world can be considered an \textit{instance} of that object. Then,
there are different instances of one and the same object at different
times and in different worlds, but it is still the same object that
is instantiated. In terms of instances, thus, the formula $p(t,w)$=``$Fo@(t,w)$''
is understood as ``the instance $o@(t,w)$ has the property $F$''.
Note that the talk of instances and the talk of objects are just two
different ways to read a proposition. We may switch back and forth
between the two readings as we like and keep in mind that doing so
does not affect the content of the proposition. There are three different
ways of using instance talk. One of these ways is the talk of \textit{time-world
instances} that we have just encountered. Another way is to translate
only the \textit{time} mode into an instance and leave the world mode
as it is. Thus, instead of speaking of an object having some property
at a specific time and in a specific world, one may speak of a \textit{time-instance}
of the object having the property in a specific world. So, for example,
instead of saying that ``Joe went to school when he was six in world
$w$'', one may say that ``six-year-old Joe went to school in world
$w$''. Formally, a time-instance of Joe is denoted as $\text{Joe}_{t}\equiv\text{Joe}@t$,
and in the given example $t$ would denote some point in time when
Joe was six years old. The third way is to translate only the world
mode into an instance, so that $\text{Joe}_{w}\equiv\text{Joe}@w$
is the instance of Joe in world $w$. 

It is helpful to visualize objects as being \textit{extended} into
both the time mode and the world mode. That is, one may conceive of
modes in a manner analogous to dimensions. Instances then collapse
objects along some of their modes. A time-instance collapses the object
along the time mode, so that it remains extended along the world mode.
For example, a time-instance of an electron is an \textit{electron
cloud}. A world-instance collapses the object along the world mode
so that it remains extended along the time mode. For example, a world-instance
of an electron is a \textit{trajectory}. Last, a time-world-instance
collapses the object along both modes, so that it loses its modal
extension and becomes a ``modal point''. For example, a time-world-instance
of an electron is a point located in 3D space. Instances inherit the
properties that the objects have at their instantiation. For example,
a time-world instance $e_{t,w}$ of an electron $e$ is located at
the position $\boldsymbol{x}$ that the electron occupies at time
$t$ in world $w$. And as for Joe, his time-world instance $\text{Joe}_{t,w}$
has the property of going to school.

It should have become clear from the setup of the framework and the
chosen formulations that I understand neither the \textit{present
time} nor the \textit{actual world} as absolute entities. Rather,
I regard all times and all worlds as equally existing, although they
do not exist in the same sense in which objects exist. An object exists
\textit{at} a time and \textit{in} a world. Without a time and a world
the object cannot exist. So times and worlds are \textit{preconditions}
of existence, or more precisely, in the terminology of the here-proposed
framework, times and worlds are \textit{modes of existence}\footnote{It should be remarked that the framework is, with regard to the theory
of relativity, a non-relativistic framework, because it treats time
not on the same footing as space. In a relativistic framework to be
developed in a future publication, time as an anchor of propositions
must be replaced by a spacelike hypersurface in four-dimensional spacetime.}.

\subsection{Particles\label{sec:Particles}}

As usual in a physical theory, we will from now on restrict the set
of objects to systems of elementary particles\textit{.} Technically,
a system of particles is a \textit{mereological sum} of particles,
that is, particles are in a part-whole relationship with systems.
A one-particle system consists of only one particle, and an $N$-particle
system consists of $N$ particles. If a particle is part of a system
which is part of another system, then the particle is also part of
the latter system. We shall not dive into the details of the theory
of mereology, and the reader may be referred to textbook literature
\citep[cf.][]{Varzi2014}. 

Systems of particles cannot be naively identified with macroscopic
objects. Macroscopic objects (ships, dogs, people), may lose or gain
particles and still remain the same macroscopic object, which is not
the case for systems. It is a highly nontrivial question how to define
macroscopic objects and how to identify them across time. Of course,
it should be no less problematic to identify macroscopic objects across
\textit{worlds}. If in one world Joe is lacking one hair compared
to our world, it would appear natural to assume that it is still \textit{Joe}
who is lacking a hair in that world, and not a completely different
person. I believe these issues can somehow be settled, so that it
makes sense to speak of one and the same object (including people)
across different times and across different worlds. Luckily, though,
these difficulties do not affect the quantum theory that I am going
to propose, as they do not arise for elementary particles. An elementary
particle at different times simply \textit{is} the same particle,
only at different times, and a particle in different worlds simply
\textit{is} the same particle, only in different worlds. (Recall that
instances are no \textit{copies}; it is \textit{the same particle}
that only occupies different positions in space at different times
and in different worlds.)

The symbol ``$o$'' denoting a system of particles in a proposition
like ``$Fo\,@(t,w)$'' is taken as a \textit{rigid designator} in
a sense analogous to how Kripke introduced the term into modal logic
\citep{Kripke1981}: the symbol refers to the same system at all times
when it exists, and in all worlds where it exists. Since we stay in
the non-relativistic domain, there is no particle creation or annihilation.
Thus, if a particle exists in one world at one time, then it exists
at all other times in that world. Let us further demand particle conservation
\textit{across worlds}, so that when a system $o$ exists at one time
and in one world, it also exists at any other time and in any other
world. There can be no particles missing and no particles being added
to the system across worlds and across time. The system of \textit{all}
particles is identified with the \textit{universe}, and the number
of particles in the universe is assumed to be finite and equal to
some fixed number $N$.

\subsection{Trajectories\label{sub:Trajectories}}

Particles have a fundamental property, which is their position in
three-dimensional space. A particle can at some time $t$ and in some
world $w$ have the position $\boldsymbol{q}$, and at another time
$t'$ and in another world $w'$ have the position $\boldsymbol{q}'$.
Let $u$ be the universe consisting of $N$ particles, and $q=(\boldsymbol{q}_{1},\ldots,\boldsymbol{q}_{N})$
be a complete list of the positions of all particles. Then this list
of positions, which is called a \textit{configuration} of the universe,
can mathematically be interpreted as a vector in the $3N$-dimensional
space $\mathcal{Q}:=\mathbb{R}^{3N}$, which is called the \textit{configuration
space}. A point in the 3$N$-dimensional configuration space corresponds
to the position of $N$ particles in the three-dimensional space (\textit{3D
space}, in short). There is a subtle but profound issue with defining
the configuration space as the vector space $\mathbb{R}^{3N}$, and
this issue will later be discussed and addressed by re-defining the
configuration space as the tensor space $\mathbb{R}^{3\times N}$
instead. The following considerations, though, are independent of
this choice of definition.

Let $F_{q}$ be the property of having the configuration $q\in\mathcal{Q}$,
then the proposition ``$F_{q}u\ @(t,w)$'' asserts that the universe
has the configuration $q$ at time $t$ and in world $w$. Now let
us assert that \textit{at every time there is for every world a unique
configuration of the universe}. In propositional calculus this is
translated into the conjunction of the following two propositions:
\begin{align}
\forall(t,w)\exists q:\quad & F_{q}u\,@(t,w)\label{eq:uniqueness1}\\
\forall(t,w,w',q,q'):\quad & \bigl(F_{q}u\,@(t,w)~\wedge~F_{q'}u\,@(t,w')\bigr)\rightarrow\bigl(q=q'~\leftrightarrow~w=w'\bigr).\label{eq:uniqueness2}
\end{align}
 The first proposition asserts that for every time and for every world
there is (at least) one configuration of the universe, and the second
proposition asserts that the configuration is unique. So there is
a set of trajectories $\Gamma=\{\gamma\}$ through configuration space,
which is in one-to-one correspondence to the set of worlds $\mathcal{W}=\{w\}$,
\begin{equation}
w~\leftrightarrow~\gamma.
\end{equation}
Consequently, we may \textit{identify} worlds and trajectories. This
should not be taken to suggest that worlds literally \textit{are }trajectories,
in the sense of a synonym. The term ``world'' just has a different
linguistic function than the term ``trajectory''. But whenever the
term ``world'' is used in the theory, we know that it \textit{corresponds}
to a unique trajectory. With regard to the considerations further
above, a trajectory is a \textit{world-instance} of the universe,
that is, it is fixed to one specific world but extended into the time
mode. A \textit{time-world-instance} of the universe would then be
a particular \textit{configuration} of the universe. 

The uniqueness relations \eqref{eq:uniqueness1} and \eqref{eq:uniqueness2}
imply that each trajectory is uniquely defined by any one of its points.
So, we may \textit{parametrize} the trajectories by their initial
configuration at $t=0$, yielding the \textit{trajectory function}
$\xi=\xi_{t}(q)$, so that each trajectory $\gamma$ is related to
the trajectory function via $\gamma_{t}=\xi_{t}(\gamma_{0})$, which
implies that $\xi_{0}(q)=q$. The trajectory function is a unique
characterization of the continuous bundle of trajectories $\Gamma$,
and hence can be regarded as an alternative representation of the
latter.

Up to now there have been no further restrictions imposed on the actual
\textit{form }of the trajectories. They are not required (so far)
to be continuous or differentiable or anything else; they might be
erratic and discontinuous, jumping around across configuration space
from one moment to another in a fractal manner, resembling a random
process rather than a deterministic evolution. The actual form of
the trajectories is what the \textit{physical laws} will yield. These
laws are sorting out unphysical trajectories from physical ones, which
leads us to a quite general definition of a physical law: a physical
law is a restriction on the set of logically possible ways the universe
may evolve in time. There are (vastly) more logically possible ways
the universe may evolve than there are physically possible ways. Consequently,
the set of \textit{logically} possible worlds is much bigger than
the set of \textit{physically} possible worlds. A physical theory
essentially defines how the set of logically possible worlds is to
be restricted to the set of physically possible worlds by conditions
imposed on the corresponding trajectories, and these conditions come
in the form of differential equations.

\section{Probability\label{sec:Probability}}

The success of quantum mechanics is often regarded as evidence for
\textit{objective probability} and \textit{objective uncertainty}.
Quantum mechanics, so we are told, enforces a picture of Nature not
being decided about ``how to proceed'' when a measurement takes
place, or even not being decided about ``how to be'' with respect
to unobserved system properties. However, Bohmian mechanics as well
as Everettian mechanics offers a different picture in this respect.
Both theories deny the necessity of objective probability. Rather,
the universe is at all times in a certain (quantum) state, and probability
only comes into play as a consequence of \textit{incomplete knowledge}
of the observer, a concept of probability that is referred to as \textit{subjective}
or \textit{epistemic probability}. In Everettian mechanics, the world
of an observer is splitting up into multiple worlds during a measurement
process, and the observer does not know which world he or she (as
a macroscopic system capable of conscious experience) will end up
in after the measurement. There are puzzling issues with this kind
of interpretation \citep[cf.][]{Kent1990,Squires1990,McInerney1991,Saunders_et_al_2008,Wallace2010},
but we will not get into detail here. In Bohmian mechanics, the particles
in the universe are at all times in a certain configuration, but the
observer does not (precisely) know which configuration that is. The
challenge of both interpretations is not merely a philosophical one,
but also how to get the empirically correct numerical values for the
probabilities as given by \textit{Born's rule}. For $\hat{\Pi}_{a}$
being a projector onto the eigenspace of some eigenvalue $a$ of an
observable $\hat{A}$, Born's rule says that for a system being at
time $t$ in a state described by the wavefunction $\Psi_{t}$, the
probability to find the observable $\hat{A}$ obtaining the value
$a$ in a measurement at time $t$, is given by
\begin{equation}
P_{t}(a)=\frac{\|\hat{\Pi}_{a}\Psi_{t}\|^{2}}{\|\Psi_{t}\|^{2}}.\label{eq:Born}
\end{equation}
As for Everettian mechanics, there are numerous attempts to derive
Born's rule using various approaches involving observer memory states
\citep{Everett1957,Everett1973}, infinite ensembles \citep{Hartle1968},
frequency operators \citep{Farhi_et_al_1989}, decoherence \citep{Saunders1998,Wallace2010},
consistent histories \citep{Omnes1992,Dowker_et_al_1996}, and decision
theory \citep{Deutsch1999,Rae2009}. All these approaches are still
controversial. As for Bohmian mechanics, Born's rule is derived on
the basis of an additional \textit{quantum equilibrium hypothesis},
whose status and justification are a controversial issue as well.
The quantum equilibrium hypothesis (QEH) asserts that the probability
density of the configuration of a system described by the wavefunction
$\Psi_{t}$ is at some time point $t$ given by
\begin{equation}
\rho_{t}(q)=\frac{|\Psi_{t}(q)|^{2}}{\int dq'\,|\Psi_{t}(q')|^{2}},\label{eq:QEH}
\end{equation}
in which case the system is said to be in \textit{quantum equilibrium.
}By virtue of the continuity equation, then, the probability distribution
$\rho_{t}$ is guaranteed to satisfy relation \eqref{eq:QEH} for
all times $t$, a feature that is called \textit{equivariance}. As
the here-proposed theory, which may be called the \textit{world continuum
theory}, is a combination of Bohmian and Everettian mechanics, it
may come to no surprise that it avoids objective probability in favor
of an epistemic account. Probability comes into play as a measure
of the subjective uncertainty of an observer about the world \textit{he
or she actually lives in}, which is a somewhat sloppy way of saying
that he or she does not precisely know which \textit{world mode} is
to be used when determining the trajectory of the universe that governs
his or her own experience. We will use that sloppy talk since it is
much shorter and simpler to grasp, but we shall keep in mind that
it has to be understood in a more sophisticated sense.

\subsection{Continuous substance}

The approach to probability that I am going to propose is based on
the concept of \textit{continuous substance}, which might appear straightforward
or even trivial to some readers, yet unfamiliar or even absurd to
others. The idea of continuous substance has somewhat come out of
fashion. Many of us have become so highly accustomed to the concept
of discrete particles, quantum jumps, and energy quanta, that they
can hardly imagine continuous substances. To them, the very concept
of measuring the \textit{quantity} of existing stuff is rigidly tied
to its discrete nature. There is, so they believe, \textit{more} stuff
of a particular kind contained within some region $X$ than there
is stuff contained within another region $X'$ if and only if there
are \textit{more particles} of that stuff in $X$ than there are in
$X'$. In contrast, children do in general not have any problems when
comparing the quantity of different heaps of continuous stuff. ``Bob
has more ice cream than me'', little Alice may be heard to complain,
and we can hardly expect her to be aware of the discrete structure
of matter. In her mind there seems to be an intuitive concept of a
``more'' relation between heaps of stuff that is not rigidly tied
to the counting of particles. To this intuitive concept I would like
to appeal.

\subsection{Substantial amount\label{sub:Substantial-amount}}

In the case of a discrete substance we can compare two heaps of substance
by counting and comparing the number of discrete constituents (particles)
of the heaps. This is not possible in the case of continuous substance.
Anyway, the notion of comparing quantities of substance should be
applicable also in this case. If there is continuous substance distributed
in space, then there should be a sense in which there is \textit{more}
substance contained within some region in space than there is contained
within another region. Being able to compare quantities of substance
is a minimal requirement to justify the notion of \textit{substance}
itself. Note that by \textit{substance} we do not necessarily mean
\textit{matter}. Substance, in the sense that we shall be using the
term here, simply means anything that is \textit{objectively existing
in physical space}. It might be matter, but it might as well be energy
or an electromagnetic field. If there were such thing as an objective
probability distribution in space, then it would constitute a \textit{substance},
too. 

Let us leave open the question whether the physical space is the traditional
$\mathbbm R^{3}$ or rather the configuration space or still some
other space. For the time being, all we require is that the physical
space $\mathcal{X}$ is a Lebesgue-measurable vector space. The Borel
sets of $\mathcal{X}$ form a $\sigma$-algebra and they are called
\textit{measurable} subsets of $\mathcal{X}$. When we speak of subsets
of $\mathcal{X}$ in the following, we shall mean \textit{measurable}
subsets.

Let there be a continuous substance $S$ distributed in space by a
density function $\rho\geq0$ over $\mathcal{X}$. We may think of
$S$ as a continuous fluid or gas. At each point $x$ in space where
$\rho(x)>0$, let us hold that there is $S$, and at those points
where $\rho(x)=0$, let us hold that there is no $S$. Moreover, if
for two points $x$ and $x'$ we have $\rho(x)=c\cdot\rho(x')$, then
we hold that S is \textit{$c$ times more densely packed} within an
infinitesimal volume centered at $x$ than it is packed at $x'$.
Put simply, $\rho$ measures not the density of a physical quantity
like mass, charge, or energy, which is associated with a physical
unit, but rather the density of the substance itself; let us, therefore,
call it the \textit{substantial density}. With $dx$ denoting the
infinitesimal volume element in $\mathcal{X}$, the integration of
$\rho$ over some finite region $X$ in space,
\begin{equation}
\mu(X)=\int_{X}dx\,\rho(x),
\end{equation}
yields the \textit{substantial amount of }$S$ contained within that
region. So, instead of asking \textit{how many} discrete constituents
of stuff are there, as we may do in the case of discrete substances,
we more generally ask \textit{how much} of the stuff is there, which
is then answered by the substantial amount. While $\mu$ is dimensionless,
the density $\rho$ has the dimension of \textit{substance per unit
volume} and its SI units are $1/m^{D}$, where $D$ is the dimension
of the physical space. We may calibrate the measure $\mu$ as we like.
For example, we may calibrate $\mu$ to a reference amount $\mu_{0}=\mu(X_{0})$
for some region $X_{0}$, such that the calibrated measure $\tilde{\mu}=\mu/\mu_{0}$
would measure the amount of substance in multiples of the reference
amount $\mu_{0}$ contained within the reference region $X_{0}$.
If the total amount of substance is finite, we may most conveniently
calibrate the measure $\mu$ to $\tilde{\mu}=\mu/\mu(\mathcal{X})$,
which means that $\tilde{\mu}$ would measure the overall \textit{proportion}
of substance contained within a given region.

The substantial density is not actually more fundamental than the
substantial amount. We could as well have started from the substantial
amount $\mu$ and then define the substantial density by
\begin{equation}
\rho(x)=\lim_{\epsilon\rightarrow0}\frac{\mu(B_{\epsilon}(x))}{\lambda(B_{\epsilon}(x))},
\end{equation}
where $B_{\epsilon}(x)$ is an $\epsilon$-ball centered at $x$,
and $\lambda$ is the Lebesgue measure on $\mathcal{X}$. Substantial
density and substantial amount are two sides of the same coin, and
this coin represents the ability to quantify the amount of substance
contained in regions of physical space.

When for two regions $X$ and $Y$ we have $\mu(X)=c\cdot\mu(Y)$,
then this means that there is \textit{$c$ times more substance contained
within $X$ than within $Y$}. It is exactly this ability to numerically
measure and compare the amount of substance contained in different
regions in space, which justifies the physical notion of ``substance''.
If different heaps of a substance could not be measured and compared
with respect to their quantity, the term ``substance'' would be
inadequate. Only, we have to refrain from the idea that a substance
must necessarily be composed of discrete entities such as particles.
It is conceivable, and mathematically describable, that a substance
is truly continuous and can still be measured with respect to its
quantity.

Note that we are \textit{not} talking about the number of mathematical
points in a set, which would be given by the \textit{cardinality}
of the set. The cardinality, or cardinal number, is an abstract mathematical
concept introduced by Georg Cantor, the inventor of set theory. Two
sets have the same cardinality exactly if there is a one-to-one mapping
between the elements of both sets. For finite sets, the cardinality
equals the number of elements in the set. For infinite sets, the cardinality
is a so-called \textit{transfinite number}, denoted by $\aleph_{0},\aleph_{1},\ldots$
, so that the set $\mathbbm N$ is defined to have the smallest transfinite
cardinality $\aleph_{0}$, which is also denoted as \textit{countably
infinite}. The cardinality $\mathfrak{c}=2^{\aleph_{0}}$ of the continuum
can be proven to be \textit{bigger} than $\aleph_{0}$ (in the sense
that there is no one-to-one mapping), but its actual value is logically
independent from the axioms of ZFC set theory. The value of $\mathfrak{c}$
depends on whether one accepts the \textit{continuum hypothesis} or
not, which postulates that $\mathfrak{c}=\aleph_{1}$. The cardinality
of sets is a profound and fruitful concept exploring the depths of
mathematical logic, but it has few to do with physical considerations.
The \textit{amount} $\mu$ of a substance contained in a given region
in space is not to be confused with the cardinal number of mathematical
points within that region. While the cardinality is the set theoretic
number of elements in a given set, $\mu$ is the \textit{integrated
spatial density} of a substance in a given region in space, and as
such it is not a property of space itself, hence not an \textit{a
priori} measure, but rather a property of the physical substance being
distributed in space, hence an \textit{a posteriori} measure. A typical
\textit{a priori} measure of space would be the \textit{spatial volume}
as provided by the Lebesgue measure. In contrast, the substantial
amount $\mu$ is an \textit{a posteriori} measure that captures the
contingent spatial distribution of a physical substance under consideration.
Different measures $\mu$ would correspond to different spatial distributions
of the substance. Yet, to speak of a substance $S$ being distributed
in space \textit{is just} to speak of a particular measure $\mu$
on that space, so that $\mu(X)$ yields the substantial amount of
$S$ contained within the region $X$. The substantial density and
accordingly the substantial amount are \textit{physical} properties
of a substance in the same manner that, say, energy is a physical
property of a system. Energy has no mathematical meaning; it only
has a mathematical \textit{form}. Same with the substantial density
and the substantial amount: they have a mathematical form, and this
form has mathematical properties. But the \textit{meaning} of these
notions is physical.

The substantial amount can be regarded as a straight generalization
of the number of particles of a discrete substance, which can be seen
by considering a substantial density of the form
\begin{equation}
\rho(x)\sim\sum_{n=1}^{N}\delta(x-x_{n}),
\end{equation}
for some finite number $N$ of point-like particles located at positions
$x_{n}$. Then, the substantial amount of stuff concentrated within
a finite region $X$ is proportional to the number of particles contained
within that region,
\begin{equation}
\mu(X)\sim\int_{X}d^{3}x\sum_{n=1}^{N}\delta(x-x_{n})=|\left\{ n\mid x_{n}\in X\right\} |.
\end{equation}
Thus, discrete substances are really just a special case where the
substantial density has singularities, which form discontinuities
of the substantial amount, and these discontinuities are identifiable
as \textit{particles} of the substance. 

If we consider the trajectories across configuration space corresponding
to different worlds as physically existing, they constitute a continuous
substance, and we can apply the above concepts, so that $\rho$ now
measures the density of trajectories in configuration space, and $\mu$
measures the amount of trajectories contained within a given region
in configuration space.

\subsection{Re-thinking Laplace\label{sub:Beyond-Laplace}}

In his famous ``Essay on Probability'', Laplace \citeyearpar{Laplace1814a,Laplace1902}
introduced probability as the degree of certainty, or credence, to
obtain a desired outcome from a finite set of possibilities. More
precisely, if $A$ is the set of favorable outcomes and $\Omega$
is the set of possible outcomes, then the probability to obtain a
favorable outcome is defined as
\begin{equation}
P(A)=\frac{|A|}{|\Omega|},\label{eq:Laplace}
\end{equation}
where $|\cdot|$ counts the number of elements. There are numerous
ways to justify Laplace's rule, but most of them are circular. For
example, deriving Laplace's rule from an assumption of uniform probability
on the set $\Omega$ only shows that Laplace's rule is \textit{consistent}
with probability theory. The very notion of probability itself, though,
cannot be derived from probability assumptions. What Laplace had in
mind was to postulate a quantity called ``probability'' that applies
to a certain kind of situation where we have to quantify our degree
of certainty that something is the case. Whenever we are completely
\textit{indifferent} which one of a given finite set of possibilities
is actually realized, then we ought to apply Laplace's rule. This
conception of probability is an \textit{epistemic} conception, that
is, it relates to the knowledge of an observer. Two observers with
distinct states of knowledge may attribute distinct probability distributions
to the same set of possibilities. There is no objective probability
distribution, so probability is nothing existing ``out there'' \citep[cf.][]{Finetti1995}.
I adhere to this conception of subjective probability, and I think
it is all one needs to also understand the probabilistic aspect of
quantum mechanics. 

The direct translation of Laplace's rule to the situation of an observer
in an objectively existing multiplicity of worlds would be to take
$\Omega$ as the set of all worlds, because each world may possibly
be the world of the observer. However, in the here-proposed theory
there is a continuum of worlds, so there is no such thing as the ``number''
of worlds, and the ratio \eqref{eq:Laplace} would be ill-defined.
We, therefore, have to generalize Laplace's rule to infinite sets,
but how may this be reasonably accomplished?

Laplace did not give any \textit{reason} why the probability should
be equal to the ratio \eqref{eq:Laplace}; he simply defined it so.
So let us fancy a justification. If $A$ and $B$ are two sets of
possibilities, and there are \textit{$c$ times more possibilities}
in $A$ than there are in $B$, then it should be \textit{$c$ times
more probable} that the actually realized possibility lies in $A$
than that it lies in $B$. Consequently, if there is a plausible measure
$\mu$ on the set $\Omega$ of possibilities, so that $\mu(A)=c\cdot\mu(B)$
means that $A$ contains $c$ times more possibilities than $B$,\textit{
}then we should expect that
\begin{equation}
\frac{P(A)}{P(B)}=\frac{\mu(A)}{\mu(B)},
\end{equation}
The final step is then a convenient normalization of probability,
$P(\Omega)=1$, so that we obtain
\begin{equation}
P(A)=\frac{\mu(A)}{\mu(\Omega)}.
\end{equation}
The advantage of these considerations is that they do not rely upon
the sets to be finite. Whenever there is a reasonable concept of an
``amount'' of possibilities provided by a measure $\mu$ on the
set of all possibilities $\Omega$, then there is a related probability
measure $P$ on $\Omega$. Of course, for different measures $\mu$
we would obtain different probability measures $P$, so there needs
to be an independent justification why a certain measure $\mu$ is
the relevant one. Such independent justification cannot be provided
by mathematics alone but has to be a \textit{physical} justification
obtained within the framework of a physical theory. The important
thing to note is that the physical theory does not have to provide
the probability concept itself. These are independent reasonings.
The physical theory yields the justification for a measure $\mu$,
and then the probability considerations apply independently. In the
finite case the measure $\mu$ is naturally provided by the number
of possibilities, but what is the relevant measure in the case of
uncountably many worlds?

In section \ref{sub:Trajectories} we have seen that there is a one-to-one
correspondence between worlds and trajectories in configuration space.
Thus, for every world $w$ there is a unique trajectory $\gamma$
in configuration space, and this trajectory is parameterized by the
time parameter $t$. The trajectory point $q=\gamma_{t}$ is then
the configuration of the universe in the world corresponding to $\gamma$
at time $t$. Different trajectories are not allowed to cross each
other, because otherwise there would be more than one world for an
individual configuration (at the crossing point of the trajectories),
which is forbidden by the uniqueness relations \eqref{eq:uniqueness1}
and \eqref{eq:uniqueness2}. A system of non-crossing trajectories
is a \textit{deterministic} system, which means that each trajectory
is completely determined by any one of its points \citep[see][for thorough discussions on the role of determinism in modern physics]{Earman2004,Vaidman2014}.
We may represent the objectively existing collection of trajectories
$\Gamma=\{\gamma\}$ by a single \textit{trajectory function }$\xi$,
so that $\gamma_{t}=\xi_{t}(\gamma_{0})$. Note that up to now we
have nowhere specified that the trajectories have to obey a differential
equation. They may be (so far) non-differentiable, yet even discontinuous,
but still they would be deterministic in the sense that they would
be completely determined by any one of their points. 

Let the substantial density of trajectories crossing an infinitesimal
region centered at $q$ be given by some density function $\rho_{t}$.
Then the substantial amount of trajectories crossing some finite region
$Q$ at time $t$, reads
\begin{equation}
\mu_{t}(Q)=\int_{Q}dq\,\rho_{t}(q).\label{eq:worldmeasure}
\end{equation}
If for two regions $X$ and $Y$ we have $\mu(X)=c\cdot\mu(Y)$, then
this means that there are $c$ times more trajectories crossing $X$
than there are trajectories crossing $Y$, where ``more'' is to
be understood in a \textit{physical }sense. Mathematically, there
are exactly as many trajectories crossing $X$ than there are trajectories
crossing $Y$, namely $\mathfrak{c}=2^{\aleph_{0}}$. But the substantial
amount is not about mathematical entities but rather about a physical
substance that is \textit{described} mathematically. There is $c$
times more \textit{substance} in $X$ than in $Y$, and this particular
substance is composed out of \textit{trajectories}. Now, since each
trajectory corresponds to exactly one world, $\mu_{t}$ measures the\textit{
substantial amount of worlds} whose trajectories cross the region
$Q$ at time $t$. Each world corresponds to a \textit{possibility},
namely the possibility that this world is \textit{our} world. When
a region $X$ contains $c$ times more worlds than another region
$Y$, then according to our probability considerations above it should
be $c$ times more probable that our world is contained in $X$ than
in $Y$, so
\begin{equation}
\frac{P_{t}(X)}{P_{t}(Y)}=\frac{\mu_{t}(X)}{\mu_{t}(Y)}.
\end{equation}
Since our world is with certainty at any time somewhere in $\mathcal{Q}$,
so $P_{t}(\mathcal{Q})=1$, it follows that
\begin{equation}
P_{t}(Q)=\frac{\mu_{t}(Q)}{\mu_{t}(\mathcal{Q})}.\label{eq:probeq}
\end{equation}
Say, Joe is at time $t$ about to measure an observable $A$ that
obtains values $a_{1},a_{2},\ldots$ in mutually disjoint and exhaustive
sets of worlds $W_{1},W_{2},\ldots$, that is, $W_{i}\cap W_{j}=\emptyset$
for $i\neq j$, and $\bigcup_{i}W_{i}=\mathcal{W}$. Furthermore,
say that the trajectories in these worlds are at time $t$ crossing
the respective regions $Q_{1},Q_{2},\ldots$ in configuration space.
Since there is a one-to-one correspondence between worlds and trajectories,
and since there is a one-to-one correspondence between trajectories
and their points at time $t$, for $t$ being arbitrarily given, the
regions $Q_{1},Q_{2},\ldots$ are also mutually disjoint and exhaustive.
Then, the probability $p_{k}$ that Joe will find himself in a world
where $A$ obtains the value $a_{k}$ is given by
\begin{equation}
p_{k}=P_{t}(Q_{k}).
\end{equation}
Since $P_{t}$ is a probability measure on $\mathcal{Q}$ and since
the regions $Q_{1},Q_{2},\ldots$ are mutually disjoint and exhaustive,
the numbers $p_{k}$ fulfill the requirements of a probability distribution,
that is $p_{k}\geq0$, and $\sum_{k}p_{k}=1$.

In the next section we have to establish a link between the substantial
density $\rho_{t}$ of trajectories in configuration space and the
wavefunction $\Psi_{t}$. This link cannot be provided other than
by \textit{postulating} that the substantial density of trajectories
crossing an infinitesimal region centered at the point $q\in\mathcal{Q}$
is given by $\rho_{t}(q)=|\Psi_{t}(q)|^{2}$. This postulate complements
another central postulate, which links the \textit{course} of each
trajectory to the wavefunction via the Bohm equation \eqref{eq:trajectoryeq}.
These two postulates give the wavefunction a \textit{physical} \textit{meaning},
namely that of a generating function for a continuous substance formed
by a bundle of trajectories distributed in configuration space. Our
world is part of this substance, traveling along its trajectory through
configuration space beneath uncountably many other worlds.

\section{The wavefunction\label{sec:Wavefunction}}

A central element of quantum theory is the wavefunction $\Psi_{t}$,
and a great challenge for any interpretation of quantum mechanics
is to give a \textit{physical meaning} to the wavefunction. Does it
represent a physical entity itself or is it just a mathematical tool
to calculate probabilities? No doubt the wavefunction is \textit{physically
significant}, as it appears in the fundamental equations from where
observable values are derived. But beyond its physical \textit{significance},
the physical \textit{meaning} of the wavefunction is subject to longstanding
controversial debates. In Bohmian mechanics, the wavefunction represents
a physical entity, also called the \textit{pilot wave} or the \textit{guiding
field}, and it is conceived of as a physical field in configuration
space that guides all particles in the universe along their trajectories.
In Everettian mechanics, the wavefunction is also a physically existing
entity, and, moreover, the wavefunction is all there is in the universe.
Particles and macroscopic objects only appear as patterns formed by
the wavefunction in the course of its temporal evolution \citep[cf.][]{Wallace2010}.
Here, I wish to propose a different picture. The wavefunction is,
in this picture, not a physically existing entity itself, but rather
an abstract mathematical tool to determine the form of a physically
existing continuum of trajectories of varying density in configuration
space, called the \textit{world continuum}, which is identified with
the history of the universe. The wavefunction thus has an \textit{ontic}
meaning rather than an \textit{epistemic} one: it represents a compact
and complete mathematical representation of the entire history of
the universe, and not of our state of knowledge \textit{about} the
universe. In the following we shall get some more into detail.

\subsection{Configuration space\label{sub:Configuration-space}}

For systems of more than one particle, the wavefunction is a function
not in the three-dimensional space but in the $3N$-dimensional configuration
space. This raw mathematical fact is a serious obstacle towards a
straightforward physical interpretation of the wavefunction. For if
the wavefunction is taken to be (or to represent) a physically existing
entity, then should not the configuration space be considered the
real space, instead of the 3D space? And even if the wavefunction
is taken to be just a mathematical construction, is the configuration
space not still a more adequate representation of physical reality
than the 3D space? 

Consider a single particle. According to the world continuum theory,
the particle is in every world and at every time located at an exact
point in 3D space. Let us concentrate on one individual world $w$
having a trajectory $\boldsymbol{\gamma}$. At each time $t$ the
particle is located at a certain position $\boldsymbol{q}=\boldsymbol{\gamma}_{t}$.
As all time points are considered to be equally real, the particle
in world $w$ is physically represented not by a single point in 3D
space but rather by its entire trajectory $\boldsymbol{\gamma}$,
which is a one-dimensional object in space and time, it is a curve
and not a point of dimension zero. As all worlds are taken to be equally
real as well, the particle is altogether physically represented by
a continuous bundle of trajectories from the set $\Gamma=\{\boldsymbol{\gamma}\}$.
In the same way as the particle's trajectory in one world is composed
out of continuously many points, one point per time point $t$, the
trajectory bundle is composed out of continuously many trajectories,
one trajectory per world. 

Now consider $N$ particles. In world $w$ and at time $t$ these
particles are located each at an exact position in 3D space. Say,
particle $n$ is located at time $t$ at the position $\boldsymbol{q}_{n}=\boldsymbol{\gamma}_{n,t}$,
with $n=1,\ldots,N$. The mathematical representation of the location
of the particles is a list of their three-dimensional positions, $q\equiv(\boldsymbol{q}_{1},\ldots,\boldsymbol{q}_{N})$.
This list of positions can mathematically be interpreted as a vector
in the $3N$-dimensional space $\mathbb{R}^{3N}$, which is commonly
taken as the configuration space. However, is the configuration space
just a mathematical construction or is it a physically existing entity?
A trajectory in configuration space can be interpreted as representing\textit{
}$N$ particles moving through the real 3D space, where ``moving''
just means that at different time points there are different positions
taken by the particles. These $N$ particles form a \textit{system},
which is the mereological sum of the particles. Thus, a trajectory
through configuration space can alternatively be interpreted as representing
one single object, the universe, moving through configuration space.
These two different interpretations are equivalent, but one interpretation,
the one that views the universe as a unified object, better reflects
the phenomenon of \textit{quantum nonlocality}. Why so? In a local
theory, each individual particle trajectory $\boldsymbol{\gamma}_{n}$
must be a solution of a hyperbolic differential equation, and indeed
classical mechanics is such a local theory (ignoring here the delicate
issue of the gravitational potential). However, quantum mechanics
is a not a local theory in this respect. The trajectories of individual
particles are \textit{not} solutions to hyperbolic differential equations.
However, the trajectory $\gamma$ of the entire universe \textit{is}
a solution to a hyperbolic differential equation, namely the Bohm
equation \eqref{eq:guidingeq}. The trajectories of individual particles
depend on the wavefunction, and the wavefunction is a function on
configuration space and not on 3D space. As a consequence, the movement
of an individual particle depends on the instantaneous positions of
the other particles, however distant in space they are. It is one
of the most puzzling features of quantum mechanics, though, that such
nonlocal interdependency between particles cannot be exploited for
superluminal signaling, as has been proven in the context of Bohmian
mechanics by Valentini \citep{Valentini1991b}. Intriguingly, thus,
quantum mechanics is a nonlocal theory, on account of the mathematical
definition of locality, but it is a local theory with respect to \textit{Einstein
locality}, which amounts to the assertion that superluminal signaling
is impossible. In other words, Nature as described by quantum mechanics
is \textit{epistemically local}, that is, it appears to observers
as local, in the sense that they cannot communicate or gain information
in a nonlocal manner, but \textit{ontologically}, with respect to
sheer existence, Nature is nonlocal. This ontological nonlocality
is not just a philosophical subtlety, but it has observable physical
implications, namely those phenomena usually termed as being typically
\textit{quantum}, such as quantum interference, tunneling, teleportation,
and the like. To better reflect the ontological nonlocality of Nature,
I consider it more adequate to view the universe as one unified entity
extending in time and configuration space, and not in time and 3D
space.

Now, there is a widely ignored problem with interpreting the configuration
space $\mathbb{R}^{3N}$ as the physical space, and it is that the
three spatial dimensions of each particle are lumped together into
one column vector with $3N$ components. While for most physicists
this might appear rather unproblematic, it makes the theory vulnerable
against a subtle but profound criticism put forward by Monton \citep{Monton2002}.
As the author writes, the problem is essentially ``\ldots{}that
nowhere in the 3N-dimensional space is it specified which dimensions
correspond to which particles'', which leads the author to conclude
that ``the wave function ontology is an undesirable ontology for
quantum mechanics''. Monton's criticism applies to those quantum
mechanical theories that entail wavefunction realism in some way,
such as Bohmian mechanics and Everettian mechanics. It would also
apply to the here-proposed theory if one favors the configuration
space as the real space in order to pronounce quantum nonlocality
as a natural phenomenon.

Monton's criticism can straightforwardly be addressed by separating
spatial dimensions and particle associations into the rank-two tensor
space $\mathbbm R^{3\times N}$, which in the following will be referred
to as the \textit{\emph{configuration space}} $\mathcal{Q}$, instead
of $\mathbb{R}^{3N}$. While a point $\boldsymbol{q}=(x,y,z)^{T}$
is a column vector in the space $\mathbbm R^{3}$, a point $q=(\boldsymbol{q}_{1},\ldots,\boldsymbol{q}_{N})$
is a rank-two tensor in the space $\mathbb{R}^{3\times N}$, that
is, a matrix 
\begin{equation}
q=\begin{pmatrix}x_{1}\cdots x_{N}\\
y_{1}\cdots y_{N}\\
z_{1}\cdots z_{N}
\end{pmatrix}.
\end{equation}
Since any tensor space is also a linear space, and in that sense still
a mathematical vector space that can be endowed with an inner product
just like ordinary rank-one vector spaces, one does not sacrifice
mathematical structure. Also, there is no difficulty in defining wavefunctions
$\Psi,\Phi$ on the tensor space $\mathbbm R^{3\times N}$ instead
of $\mathbbm R^{3N}$, so that the Hilbert inner product can be defined
as 
\begin{equation}
\langle\Psi|\Phi\rangle:=\int dq\,\Psi^{*}\Phi,
\end{equation}
which is a shorthand notation of 
\begin{align}
\int dq\,\Psi^{*}\Phi & \equiv\int_{\mathbb{R}^{3}}d^{3}q_{1}\cdots\int_{\mathbb{R}^{3}}d^{3}q_{N}\,\Psi^{*}(\boldsymbol{q}_{1},\ldots,\boldsymbol{q}_{k})\Phi(\boldsymbol{q}_{1},\ldots,\boldsymbol{q}_{k}).
\end{align}
A trajectory $\gamma$ represents the movement of $N$ particles through
3D space in time. A point $q=\gamma_{t}$ on the trajectory represents
the configuration of the universe at time $t$, and the $(i,j)$-th
component $q_{ij}=(\boldsymbol{q}_{j})_{i}$ corresponds to the $i$-th
spatial component of the position of the $j$-th particle at $t$.
The confusion of spatial dimensions and particle associations criticized
by Monton does not arise, as it is clearly specified which spatial
dimensions correspond to which particles.

\subsection{World continuum\label{sub:Metaworld}}

\begin{figure}
\begin{centering}
\includegraphics{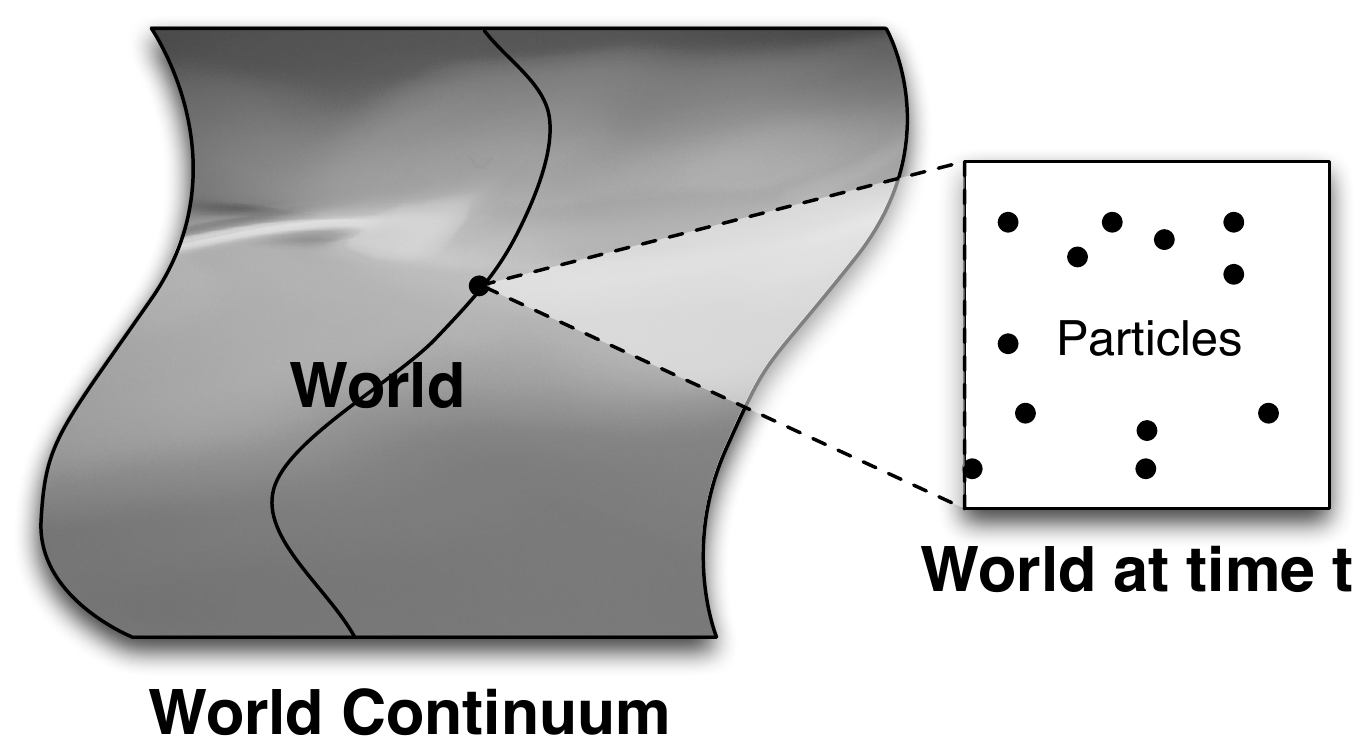}
\par\end{centering}

\caption{\label{fig:metaworld}Schematic illustration of the ontology of the
proposed theory. The world continuum is a continuous bundle of trajectories
in configuration space, with each trajectory uniquely corresponding
to a world. The density of the trajectories in configuration space
is defined by a density function $\rho_{t}=|\Psi_{t}|^{2}$. Each
trajectory $\gamma$ represents the movement of $N$ point-like particles
through 3D space and is a solution of the Bohm equation $d\gamma_{t}/dt=j_{t}/\rho_{t}$,
where the flow $j_{t}$ is also derived from the wavefunction. From
the density $\rho_{t}$, the Born rule can be derived as the subjective
probability of finding oneself within a particular world. The wavefunction
itself is not a physically existing entity but merely serves as a
generating function for the density and the flow that both determine
the form of the physically existing world continuum.}
\end{figure}

According to the view so far developed, the universe as a whole is
physically represented by a continuous bundle $\Gamma=\{\gamma\}$
of trajectories in configuration space, where each trajectory $\gamma\equiv(\boldsymbol{\gamma}_{1},\ldots,\boldsymbol{\gamma}_{N})$
represents the movement of all particles in the universe. The entire
trajectory bundle $\Gamma$ can be represented by a single trajectory
function $\xi=\xi_{t}(q)$, so that for each $\gamma\in\Gamma$ we
have $\gamma_{t}=\xi_{t}(\gamma_{0})$. In order for the trajectory
bundle $\Gamma$ to be considered as a continuous substance, the \textit{world
continuum, }there is one more feature to be provided: the \textit{substantial
density} of the trajectories in configuration space. At any given
time $t$, the trajectories may in some regions of configuration space
be more densely packed than in other regions, and this feature is
governed by a time-dependent density $\rho=\rho_{t}(q)$ on configuration
space, which is linked to a the measure $\mu=\mu_{t}(Q)$ on configuration
space via \eqref{eq:worldmeasure}. So, the world continuum is uniquely
represented by the tuple $\Upsilon=(\rho,\xi)$. The world continuum,
so I propose, is an adequate and complete picture of the history of
the physically existing universe. The wavefunction $\Psi$ is not
a physically existing entity itself but rather an abstract \textit{generating
function} that determines the \textit{form} of the world continuum.
In order for this to make sense, the wavefunction $\Psi$ must contain
all information necessary to uniquely determine the form of the world
continuum, that is, the course of each individual trajectory as well
as the density of the trajectories in configuration space. The course
of each individual trajectory is determined by the \textit{guiding
equation} of Bohmian mechanics, which in the context of the here-proposed
theory should be rather called a \textit{trajectory equation}, or
simply the \textit{Bohm equation. }That is, each trajectory $\gamma$
is a solution of the first-order differential equation
\begin{equation}
\frac{d}{dt}\gamma_{t}=\frac{j_{t}}{\rho_{t}},\label{eq:guidingeq}
\end{equation}
where the vector field 
\begin{equation}
j_{t}\equiv(\boldsymbol{j}_{t,1},\ldots,\boldsymbol{j}_{t,N})\label{eq:floweqtotal}
\end{equation}
is called the \textit{world flow}, defined by
\begin{equation}
\boldsymbol{j}_{t,n}=\frac{\hbar}{2m_{n}i}\left(\Psi_{t}^{*}\boldsymbol{\nabla}_{n}\Psi_{t}-\Psi_{t}\boldsymbol{\nabla}_{n}\Psi_{t}^{*}\right)\label{eq:floweq}
\end{equation}
and where the scalar field
\begin{equation}
\rho_{t}=|\Psi_{t}|^{2}\label{eq:densityeq}
\end{equation}
is called the \textit{world density}. The set of all solutions yields
the set $\Gamma=\{\gamma\}$, and each trajectory $\gamma\in\Gamma$
corresponds to exactly one world $w\in\mathcal{W}$. Since the set
of solutions is uncountable, so is the set of worlds. 

Now, why is \textit{$\rho$} called the \textit{world density} and
$j$ called the \textit{world flow}? Because if the wavefunction $\Psi$
obeys the \textit{Schrödinger equation}
\begin{equation}
i\hbar\frac{\partial}{\partial t}\Psi_{t}=\hat{H}\Psi_{t}
\end{equation}
for some given Hamiltonian $\hat{H}$, then it can easily be shown
that the functions $\rho$ and $j$ fulfill the \textit{continuity
equation}
\begin{equation}
\frac{\partial}{\partial t}\rho_{t}+\nabla\cdot j_{t}=0,\label{eq:continuity}
\end{equation}
where $\nabla\equiv(\boldsymbol{\nabla}_{1},\ldots,\boldsymbol{\nabla}_{N})$
is the nabla operator on configuration space, and where the scalar
product of two vectors $a,b\in\mathcal{Q}$ is defined as $a\cdot b:=\sum_{n=1}^{N}\boldsymbol{a}_{n}\cdot\boldsymbol{b}_{n}$.
As \citet{Madelung1927} already pointed out, the functions $\rho$
and $j$ describe a locally conserved compressible fluid in configuration
space. In fact, Bohmian trajectories are nothing but the \textit{pathlines}
of this fluid. However, Madelung could not provide a consistent physical
interpretation of this mathematical fact, so the hydrodynamical interpretation
of quantum mechanics was abandoned. Recently, the hydrodynamic interpretation
experienced a renaissance, and it was shown that the wavefunction
can be completely removed from the theory, leaving only trajectories
as the physically existing objects from where all observable values
can be calculated \citep{Holland2005,Poirier2010,Schiff_et_al_2012}.
However, these approaches leave it open as to how the fluid is interpreted
physically. \citet{Holland2005} remarks that their model is particularly
suited to interpret the fluid as being composed of a continuum of
``probability elements'', in line with the standard view of interpreting
the continuity equation as describing a ``probability fluid''. Whatever
probability really is, it is certainly not a material entity. In the
interpretation of the here-proposed theory, in contrast, the fluid
is interpreted as a material entity, the world continuum, and the
fluid elements flowing through configuration space are \textit{worlds}\footnote{To say that a world \textit{flows} through configuration space is
another way of saying that the configuration of the universe in that
world is parameterized by time.}. 

Under the usual assumption that the wavefunction vanishes rapidly
enough at infinity, one may integrate the continuity equation over
the configuration space, which yields
\begin{equation}
\frac{\partial}{\partial t}\int dq\,\rho_{t}=0.
\end{equation}
Thus the integral of the world density $\rho_{t}$ over the entire
configuration space is constant in time and we have
\begin{equation}
\int dq\,\rho_{t}=\int dq\,\rho_{t'}=\mu_{0},
\end{equation}
for any two time points $t,t'$. We then define the \textit{world
amount} by
\begin{equation}
\mu_{t}(Q)=\int_{Q}dq\,\rho_{t},\label{eq:volumeeq}
\end{equation}
so the total amount of worlds $\mu_{t}(\mathcal{Q})$ is constant
in time and equal to $\mu_{0}$. Physically, the constancy of $\mu_{t}(\mathcal{Q})$
means that there are no worlds being destroyed or created during the
evolution of the universe. As each world corresponds to a really existing
trajectory in configuration space, the density function $\rho_{t}$
is not a probability density but the \textit{substantial density}
of really existing trajectories. That is, the integral of $\rho_{t}$
over some region $Q$ in configuration space does not yield a probability,
but rather the \textit{substantial amount} $\mu_{t}(Q)$ of trajectories
crossing the region $Q$ at time $t$. Due to local conservation of
$\mu$, as expressed by the continuity equation \eqref{eq:continuity},
the trajectories have no beginnings and no endings, and neither do
they split nor converge. This stands in contrast to Everettian mechanics
with its ontology of splitting worlds. As we have seen earlier, one
can derive from $\mu_{t}$ the epistemic probability $P_{t}$ corresponding
to the ignorance of an observer about which world it is that he or
she lives in. So, importantly, probability is in the world continuum
theory a derived concept rather than a fundamental one. In a continuum
of worlds distributed with a certain density $\rho_{t}$, the subjective
probability to find oneself within a particular world \textit{must
}be given by \eqref{eq:probeq}, as a result of probability considerations
external to, and independent from, the physical theory itself.

The wavefunction as a generating function of the world continuum contains
two redundant parameters, which is the \textit{global scale} and the
\textit{global phase}. This is because the Bohm equation \eqref{eq:guidingeq}
is symmetric under the transformation $\Psi_{t}\mapsto re^{i\phi}\Psi_{t}$
for $r>0$ and $0\leq\phi<2\pi$, so the trajectories do not depend
on the global scale and the global phase. The world density \eqref{eq:densityeq}
is symmetric with respect to the global phase $\phi$ but it scales
quadratically with the global scale $r$. However, a global scaling
of the world density bears no physical significance, because it leaves
the relative proportions between different world amounts untouched.
A region $X$ will still contain $c$ times more worlds than another
region $Y$, irrespective of the global scaling factor. 

Besides the global scaling and the global phase there are no further
redundancies, as can be seen by writing the wavefunction in polar
decomposition $\Psi_{t}=R_{t}e^{iS_{t}}$, so that the Bohm equation
\eqref{eq:guidingeq} and the world density \eqref{eq:densityeq},
respectively, become
\begin{equation}
\frac{d}{dt}\boldsymbol{\gamma}_{n}=\frac{\hbar}{m_{n}}\boldsymbol{\nabla}_{n}S_{t},
\end{equation}
and
\begin{equation}
\rho_{t}(q)=R_{t}^{2}.
\end{equation}
Thus, the phase $S_{t}$ of the wavefunction generates the trajectory
bundle $\Gamma=\{\gamma\}$ represented by the trajectory function
$\xi$, and the amplitude $R_{t}$ generates the trajectory density
$\rho$. Altogether the wavefunction can be regarded as a generating
function of the tuple $\Upsilon=(\rho,\xi)$ that mathematically represents
the world continuum (Figure \ref{fig:metaworld}). The wavefunction
(not the projective ray in Hilbert space, of which the wavefunction
is a representative) contains slightly more information than $\Upsilon$.
If from both the wavefunction and from the world continuum $\Upsilon$
all observable values can be calculated, then the world continuum
is a slightly \textit{less }redundant representation of physical reality
than the wavefunction. The world continuum is, however, just as informative
as a projective ray in Hilbert space, which is an equivalence class
of wavefunctions differing only by their global scale and phase. From
the perspective of Occam's razor, thus, the world continuum theory
is just as ontologically demanding as any theory that entails wavefunction
realism. This includes Bohmian mechanics as well as Everettian mechanics.
As for Bohmian mechanics, the ontological costs are somewhat higher,
because in addition to a physically existing wavefunction there are
$N$ physically existing point-like particles. As for the Copenhagen
interpretation, for that matter, the wavefunction is usually not taken
as a physically existing entity but rather as a mathematical tool
to calculate probabilities. However, the Copenhagen interpretation
has issues of its own, most of which are comprised under the term
``measurement problem'', but these are not under discussion here.

\section{Foundations of the theory\label{sec:Foundations}}

From the preceding considerations we shall now distill a minimal set
of postulates that generates the theory. For the sake of simplicity,
we will stay with the case of spin-free particles. Spin can be included
in the same manner as in Bohmian mechanics by promoting the scalar
wavefunction $\Psi=\Psi_{t}(q)$ to a spinor wavefunction $\vec{\Psi}=\Psi_{t}^{\sigma_{1}\cdots\sigma_{N}}(q)$,
and by extending the Hamiltonian to include spin interaction \citep[cf.][]{Oriols2012}.

\subsection{Axiomatics}

\paragraph{Postulate 1}

The physical history of a closed system of $N$ spin-free particles
is completely described by a \textit{wavefunction} $\Psi=\Psi_{t}(q)$.
Each time-instance $\Psi_{t}$ is a vector in the Hilbert space $\mathcal{H}=L^{2}(\mathbbm R^{3\times N})$,
and is called the \textit{state} of the system at time $t$. The wavefunction
$\Psi$ is a solution of the \textit{Schrödinger equation}
\begin{equation}
i\frac{d}{dt}\Psi_{t}=\hat{H}\Psi_{t},\label{eq:schroedinger}
\end{equation}
where $\hat{H}$ is the Hamiltonian of the system, given by
\begin{equation}
\hat{H}=\sum_{n=1}^{N}\frac{\hat{\boldsymbol{p}}_{n}^{2}}{2m_{n}}+V(\hat{q}),
\end{equation}
where $\hat{\boldsymbol{p}}_{n}=\frac{\hbar}{i}\boldsymbol{\nabla}_{n}$
is the momentum operator of the $n$-th particle, and $\hat{\boldsymbol{q}}_{n}\Psi(q)=\boldsymbol{q}_{n}\Psi(q)$
is the position operator of the $n$-th particle, so that $\hat{q}=(\hat{\boldsymbol{q}}_{1},\ldots,\hat{\boldsymbol{q}}_{N})$
is the configuration operator, and the operator-valued function $V(\hat{q})$
is the potential energy of the system.

\paragraph{Postulate 2}

The physical history of the system in a specific \textit{world }is
completely described by a trajectory $\gamma=\gamma_{t}(q)$, which
is a solution of the \textit{Bohm equation}
\begin{equation}
\frac{d}{dt}\gamma_{t}=\frac{j_{t}}{\rho_{t}},\label{eq:trajectoryeq}
\end{equation}
 where $j_{t}=(\boldsymbol{j}_{1,t},\ldots,\boldsymbol{j}_{N,t})$
is defined by
\begin{equation}
\boldsymbol{j}_{n,t}=\frac{\hbar}{2m_{n}i}\left(\Psi_{t}^{*}\boldsymbol{\nabla}_{n}\Psi_{t}-\Psi_{t}\boldsymbol{\nabla}_{n}\Psi_{t}^{*}\right),\label{eq:j}
\end{equation}
and where $\rho_{t}$ is defined by
\begin{equation}
\rho_{t}=|\Psi_{t}|^{2}.\label{eq:rho}
\end{equation}
Each solution $\gamma$ of the Bohm equation \eqref{eq:trajectoryeq}
is a \textit{world-instance} of the system, and the world-time-instance
$\gamma_{t}$ is the \textit{configuration} of the system at time
$t$ in the world corresponding to $\gamma$.

\paragraph{Postulate 3}

The trajectories form a \textit{continuous substance }in configuration
space, with the function $\rho=\rho_{t}(q)$ representing the \textit{substantial
density} of trajectories. Therefore, the \textit{substantial amount}
of trajectories crossing a finite region $Q$ in configuration space
at time $t$ is given by

\begin{equation}
\mu_{t}(Q)=\int_{Q}dq\,\rho_{t}(q),\label{eq:mu}
\end{equation}
so that for two regions $X$ and $Y$ in configuration space with
$\mu_{t}(X)=c\cdot\mu_{t}(Y)$ this means that there are $c$ times
more trajectories contained within $X$ than there are contained within
$Y$. As each trajectory corresponds to exactly one world, $\rho$
is also called the \textit{world density} and $\mu_{t}$ is also called
the \textit{world amount}.

\subsection{Remarks}

\paragraph{Remark 1}

The theory provides the following picture of reality, for a given
closed system $S$ that can also represent the entire universe: 
\begin{align}
\text{wavefunction}\quad\Psi & \quad\leftrightarrow\quad\text{physical history of \ensuremath{S}}\\
\text{state}\quad\Psi_{t} & \quad\leftrightarrow\quad\text{time-instance of \ensuremath{S}}\\
\text{trajectory}\quad\gamma & \quad\leftrightarrow\quad\text{world-instance of \ensuremath{S} }\\
\text{configuration}\quad\gamma_{t} & \quad\leftrightarrow\quad\text{world-time-instance of \ensuremath{S}}
\end{align}

\paragraph{Remark 2}

Since \eqref{eq:schroedinger} and \eqref{eq:trajectoryeq} are first-order
hyperbolic equations, they have a unique solution for every valid
initial condition. More precisely, for every well-behaved initial
quantum state $\Psi_{0}\in\mathcal{H}$ at time $t=0$ there is a
unique wavefunction $\Psi$ that is obtained by applying the unitary
time evolution operator
\begin{equation}
\hat{U}(t)=e^{-i\hat{H}t},\label{eq:U}
\end{equation}
so that
\begin{equation}
\Psi_{t}=\hat{U}(t)\Psi_{0}.\label{eq:Uinitial}
\end{equation}
Similarly, for every initial configuration $\gamma_{0}\in\mathcal{Q}$
for which $\rho_{0}(\gamma_{0})>0$, there is a unique trajectory
$\gamma$ obtained by applying the \textit{trajectory function
\begin{equation}
\xi_{t}=\int_{0}^{t}dt'\,\frac{j_{t'}}{\rho_{t'}},\label{eq:trajectoryfunction}
\end{equation}
}so that 
\begin{equation}
\gamma_{t}=\xi_{t}(\gamma_{0}).
\end{equation}
Hence, in each world there is a concrete path through space that the
particles take, and which is determined by the trajectory function
$\xi_{t}$ applied to the initial configuration $\gamma_{0}$ at $t=0$.
The path $\boldsymbol{x}_{n}(t)$ of an individual particle $n$ can
be extracted from the trajectory function $\xi_{t}=(\boldsymbol{\xi}_{1,t},\ldots,\boldsymbol{\xi}_{N,t})$
by fetching the components corresponding to that particle, so that
$\boldsymbol{x}_{n}(t)=\boldsymbol{\xi}_{n,t}(\gamma_{0})$. Being
a time-dependent function on the configuration space, the trajectory
function $\xi_{t}$ can also be regarded as a \textit{dynamical vector
field} in configuration space, assigning each point $q\in\mathcal{Q}$
its time-evolved counterpart $q_{t}=\xi_{t}(q)$. Using the trajectory
function, the time-evolved world density can also be written as
\begin{equation}
\rho_{t}(q)=\int dq'\,\delta(q-\xi_{t}(q'))\rho_{0}(q').
\end{equation}

\paragraph{Remark 3}

Due to the Schrödinger dynamics, the functions $\rho$ and $j$ can
be shown to obey the \textit{continuity equation}
\begin{equation}
\frac{d}{dt}\rho_{t}+\nabla\cdot j_{t}=0,\label{eq:conteq}
\end{equation}
so that $j$ takes the role of a \textit{trajectory current}. The
\textit{substantial flow} of trajectories crossing at time $t$ a
$(2\times N)$-dimensional directed submanifold $F=F_{1}\times\cdots\times F_{N}$,
where each $F_{n}$ is a two-dimensional directed submanifold in $\mathbbm R^{3}$,
is given by 
\begin{equation}
\nu_{t}(F)=\int_{F}dF\cdot j_{t},
\end{equation}
where $dF=(d\boldsymbol{F}_{1},\ldots,d\boldsymbol{F}_{N})$ is the
infinitesimal normal vector element on the directed submanifold $F$.
As each trajectory corresponds to exactly one world, $j$ is also
called the \textit{world current}. Negative values for the substantial
flow indicate that there are more worlds with particles crossing the
directed submanifold $F$ in reverse direction.

\subsection{Measurement\label{sec:Measurement}}

In standard quantum mechanics, measurement is an additional concept
different from that of ordinary Schrödinger evolution. In the world
continuum theory, just like in Bohmian mechanics and Everettian mechanics,
measurement is a specially designed but otherwise ordinary physical
process that involves a short and strong interaction between the system
of interest and a macroscopic measurement device involving a large
number of particles. The measurement process is modeled in the same
way as it is done in Bohmian mechanics, and in a similar way as in
Everettian mechanics, except that the ``pointer basis problem''
of Everettian mechanics does not show up, because we can adopt from
Bohmian mechanics the concept of pointer states as spatially separated
wave packets. The key idea is that different sets of pointer configurations
correspond to different measurement outcomes, so when a pointer configuration
lies within some region $Z_{a}$ of the pointer configuration space,
then this is taken to indicate the measurement outcome ``$a$''.
In contrast to Bohmian mechanics, though, there is no ``true'' configuration,
but rather all configurations equally exist in different worlds.

Consider a collection of mutually disjoint compact regions $Z_{a}$
in the pointer configuration space $\mathcal{Q}_{M}$, so that for
$a\neq a'$ we have 
\begin{equation}
Z_{a}\cap Z_{a'}=\emptyset.\label{eq:Zdisjoint}
\end{equation}
Furthermore, let there be a corresponding collection of pointer states
$\eta_{a}$, which have almost all of their support within the corresponding
regions $Z_{a}$, so
\begin{equation}
\int_{Z_{a}}dz\,|\eta_{a}(z)|^{2}\approx1.\label{eq:quasisupp}
\end{equation}
Let us call $Z_{a}$ an \textit{effective support} of $\eta_{a}$.
From \eqref{eq:Zdisjoint} and \eqref{eq:quasisupp}, and since the
pointer states are normalized to unity, it follows that the pointer
states have almost no spatial overlap, so for $a\neq a'$ we have
\begin{equation}
\eta_{a}(z)\eta_{a'}(z)\approx0,\label{eq:nooverlap}
\end{equation}
so the pointer states are \textit{quasi-orthogonal} to each other.
In the idealized case where the overlap of the pointer states is exactly
zero, the pointer states are perfectly orthogonal to each other. However,
this would require each pointer state to have a compact support in
configuration space, and such wavefunctions typically have infinite
average kinetic energy due to discontinuities of the wavefunctions
and their derivative at the boundary, which is clearly an unrealistic
scenario. Thus, the pointer states must have infinite support, and
so their mutual overlap cannot be zero. It can be made sufficiently
small, however, so that the state are at least quasi-orthogonal to
each other, which suffices to reproduce the predictions of standard
quantum mechanics to a degree that is only limited by the technological
state of the art.

The nature of the measurement interaction is such that during a short
measurement period $T_{M}$, the system of interest $S$ is coupled
to the measurement device $M$ by a strong interaction $\hat{W}_{SM}$,
so that the unperturbed Hamiltonian can be neglected,
\begin{equation}
\hat{H}_{S}+\hat{H}_{M}+\hat{W}_{SM}\approx\hat{W}_{SM}.\label{eq:HW}
\end{equation}
Before and after the measurement period, the interaction term is zero,
so that the system of interest and the measurement device evolve independently
from each other. The shortness of the measurement period $T_{M}$
can be more precisely defined by the requirement that the free evolution
of the system of interest during a period of length $T_{M}$ can be
neglected, that is
\begin{equation}
e^{-i\hat{H}_{S}T_{M}}\approx\mathbbm1.\label{eq:tmshort}
\end{equation}

During a measurement of some observable $\hat{A}=\sum_{a}a\,\hat{\Pi}_{a}$,
the system of interest becomes entangled with the measurement device,
so that each pointer state $\eta_{a}$ becomes correlated with the
projection of the wavefunction $\psi$ of the system of interest on
the subspace corresponding to the eigenvalue ``$a$'', thus
\begin{equation}
\psi\otimes\eta_{R}\quad\rightarrow\quad\sum_{a}\hat{\Pi}_{a}\psi\otimes\eta_{a},\label{measurement}
\end{equation}
where $\eta_{R}$ is the ``ready'' state of the measurement device. 

As an example \citep{Neumann1932,Bohm1952a,Everett1957}, the interaction
term may be given by 
\begin{equation}
\hat{W}_{SM}=-g(\hat{A}\otimes\hat{p}_{z}),
\end{equation}
where $g$ is a sufficiently large coupling constant and $\hat{p}_{z}$
is the momentum operator conjugate to the configuration operator $\hat{z}$
of the measurement device. Because of \eqref{eq:HW} the state of
the total system after measurement reads
\begin{align}
e^{ig(\hat{A}\otimes\hat{p}_{z})T_{M}}\left[\psi\otimes\eta_{R}\right] & =\sum_{a}\hat{\Pi}_{a}\psi\otimes e^{igaT_{M}\hat{p}_{z}}\eta_{R}\\
 & =\sum_{a}\psi_{a}\otimes\eta_{a},
\end{align}
where the functions
\begin{equation}
\eta_{a}(z)=\eta_{R}(z-gaT_{M})
\end{equation}
each have an effective support given by
\begin{equation}
Z_{a}=\{z\mid z-gaT_{M}\in Z\},
\end{equation}
where $Z$ is an effective support of the ``ready'' state $\eta_{R}$.
Depending on the measurement duration $T_{M}$ and on the separation
of the eigenvalues $a$, the coupling constant $g$ must be chosen
large enough, so that condition \eqref{eq:nooverlap} is met. 

Directly after measurement the wavefunction $\Psi'$ of the total
system is a sum of branches,
\begin{equation}
\Psi'=\sum_{a}\Psi'_{a},\label{eq:sumbranch}
\end{equation}
with each branch
\begin{equation}
\Psi'_{a}=\hat{\Pi}_{a}\psi\otimes\eta_{a}\label{eq:branch}
\end{equation}
representing a different measurement outcome corresponding to the
eigenvalue ``$a$'' of the observable $\hat{A}$. Because of \eqref{eq:nooverlap}
the branches are approximately orthogonal, so we have
\begin{equation}
|\Psi'|^{2}\approx\sum_{a}|\Psi_{a}'|^{2}.\label{eq:Psisquare}
\end{equation}
The measurement has the outcome ``$a$'' in world $w$ when the
configuration of that world lies within the region $Q_{a}=\mathcal{Q}_{S}\times Z_{a}$,
which is a formal way of stating that in world $w$ the pointer shows
a value corresponding to the outcome ``$a$''. So far the objective
description. In order to see what happens in a particular world, we
have to go to the subjective description from the perspective of a
particular observer Joe. Denote the world-instance of observer Joe
in world $\overline{w}$ by $\overline{{\rm Joe}}\equiv Joe@\overline{w}$,
and denote by $\overline{q}$ the configuration of world $\overline{w}$
right after the measurement. By construction, $\overline{{\rm Joe}}$
will read off the measurement result ``$a$'' if and only if he
finds the pointer in one of the configurations contained in $Z_{a}$,
which means for the total system that the configuration $\overline{q}$
of $\overline{{\rm Joe}}$'s world $\overline{w}$ must be contained
in the region $Q_{a}=\mathcal{Q}_{S}\times Z_{a}$. The probability
for this to happen reads according to~\eqref{eq:probeq} 
\begin{align}
P(\overline{q}\in\mathcal{Q}_{a}) & =\frac{\mu(Q_{a})}{\mu(\mathcal{Q})}=\frac{\int_{Q_{a}}dq\,|\Psi'(q)|^{2}}{\int dq\,|\Psi'(q)|^{2}}\label{proba}\\
 & \approx\frac{\int_{\mathcal{Q_{S}}}dx\int_{Z_{a}}dz\,\sum_{a'}|\hat{\Pi}_{a'}\psi(x)|^{2}|\eta_{a'}(z)|^{2}}{\|\Psi\|^{2}}\\
 & \approx\frac{\int_{\mathcal{Q}_{S}}dx\,|\hat{\Pi}_{a}\psi(x)|^{2}}{\|\Psi\|^{2}}\\
 & =\frac{\|\hat{\Pi}_{a}\Psi\|^{2}}{\|\Psi\|^{2}},
\end{align}
where we have used \eqref{eq:Psisquare} and \eqref{eq:quasisupp}.
So the probability that $\overline{{\rm Joe}}$ obtains the measurement
result ``$a$'' approximately coincides with the probability given
by the Born rule~\eqref{eq:Born}. The degree of the approximation
depends on the spatial separation of the pointer wave packets after
measurement, which in turn depends on the strength and duration of
the measurement interaction.

Let us go further and derive the ``collapse of the wavefunction'',
which here becomes a merely\emph{ subjective} collapse experienced
in each world separately and differently. Let $t_{M}$ be the time
immediately after the measurement is finished, so that the post-measurement
wavefunction is given by $\Psi_{t_{M}}$. For $t>t_{M}$ the wavefunction
will evolve according to
\begin{equation}
\Psi_{t}=\hat{U}(t-t_{M})\Psi_{t_{M}}.
\end{equation}
With $\overline{\gamma}$ being the trajectory of the universe in
world $w$, the post-measurement configuration of $\overline{{\rm Joe}}$'s
world reads $\overline{q}=\overline{\gamma}_{t_{M}}$. From the moment
right after the measurement, the trajectory of $\overline{{\rm Joe}}$'s
world will evolve according to~\eqref{eq:trajectoryeq}, with the
point $\overline{q}\in Q_{a}$ fixing which trajectory is his one.
Since the Bohm equation \eqref{eq:guidingeq} is hyperbolic, the further
course of $\overline{{\rm Joe}}$'s world for $t>t_{M}$ depends on
the new ``initial'' configuration $\overline{q}$. By construction
$\overline{q}$ is somewhere in $Q_{a}$ at time $t_{M}$, and for
any $\overline{q}\equiv(\overline{x},\overline{z})$ in $Q_{a}$,
we have $\overline{x}\in\mathcal{Q}_{S}$ and $\overline{z}\in Q_{a}$,
and so the wavefunction of $\Psi_{t_{M}}$ evaluated at $\overline{q}$
reads 
\begin{align}
\Psi_{t_{M}}(\overline{q}) & =\sum_{a'}\hat{\Pi}_{a'}\psi(\overline{x})\eta_{a'}(\overline{z})\\
 & \approx\hat{\Pi}_{a}\psi(\overline{x})\eta_{a}(\overline{z})\\
 & =(\hat{\Pi}_{a}\otimes\mathbbm1)\Psi_{t_{M}}(\overline{q})
\end{align}
where we have used that $\eta_{a'}(\overline{z})\approx0$ for $a'\neq a$.
Thus, from the moment right after the measurement, the wavefunction
that governs the future fate of $\overline{{\rm Joe}}$'s world becomes
\textit{subjectively} equal to the collapsed wavefunction $\overline{\Psi}_{a}:=(\hat{\Pi}_{a}\otimes\mathbbm1)\Psi_{t_{M}}$,
although the wavefunction is \emph{objectively} uncollapsed. From
now on, since the Schrödinger equation is linear, the future fate
of $\overline{{\rm Joe}}$'s world is subjectively governed by the
time-evolved collapsed wavefunction 
\begin{equation}
\overline{\Psi}_{a,t}=\hat{U}(t-t_{M})\overline{\Psi}_{a},
\end{equation}
where $t>t_{M}$. In contrast to Everettian mechanics, there is\textit{
}\textit{\emph{no splitting of worlds}}. Before and after the measurement
the total amount of worlds is the same and given by $\mu_{t}(\mathcal{Q})$,
no worlds are being created or destroyed, split or combined. What
happens is that due to the measurement process, the configuration
space is \emph{partitioned} into smaller volumes that contain those
worlds where the individual measurement outcomes occur. As the theory
is deterministic at the level of individual worlds, which follows
from the unique solvability of the Bohm equation~\eqref{eq:trajectoryeq},
the measurement result obtained in each individual world is determined
from the very beginning (see \citealp{Vaidman2014} for a similar
view on determinism in quantum mechanics, including a critical review
on the ideas proposed in \citealp{Bostrom2012}). It only \emph{appears}
to be random to the individual observer who spends their lifetime
in a particular trajectory without knowing which one. The conundrum
of the splitting of persons that occurs in Everett's theory does not
show up \citep[see][for an intriguing analysis making the splitting of persons less bizarre]{Saunders_et_al_2008}.

Concluding, the here-proposed theory explains (1) the subjective occurrence
of probabilities, (2) their quantitative value as given by the Born
rule, and (3) the apparently random ``collapse of the wavefunction''
caused by the measurement process and by the subjective experience
of individual observers, while remaining an objectively deterministic
theory.

\section{Discussion\label{sec:Discussion}}

Let me discuss certain aspects of my approach, in particular with
respect to theories, ideas and concepts found in the literature, which
are more or less close to the ideas put forward here. The following
topics are not ordered by their importance or by the similarity of
the discussed concepts to my approach, but rather the ordering tries
to follow a thematic thread.

\subsection{Bohmian mechanics\label{sub:Bohmian-mechanics}}

Although the here-presented theory may be considered a variant of
Bohmian mechanics \citep[see][for excellent reviews]{Durr_et_al_1992,Deotto_et_al_1998,Nikolic2007,Goldstein2009,Oriols2012},
there are conceptual differences between the two theories, in particular
with regard to certain critical issues.

The first issue is related to the empirically undeniable statistical
character of the measurement results. Just like Everettian mechanics,
Bohmian mechanics is a \emph{deterministic theory}, and there seems
to be \emph{prima facie} no reason why the particles occupy one branch
of the post-measurement state rather than another, with a probability
whose value is precisely given by Born's rule~\eqref{eq:Born}. The
proponents of Bohmian mechanics argue that the probability that appears
in Born's rule is an \textit{\emph{epistemic}} quantity related to
the ignorance of the observer concerning the \textit{\emph{initial
particle configuration}}. In analogy to classical statistical mechanics,
so their argument, one must then introduce a \textit{probability density}
$\rho$ on configuration space that captures the ignorance about the
\textit{actual} configuration which is a result of the ignorance about
the initial configuration. The predictions of Bohmian mechanics are
indistinguishable from those of conventional quantum mechanics, exactly
if 
\begin{equation}
\rho_{t}=|\Psi_{t}|^{2}\label{equi}
\end{equation}
for some arbitrary initial time $t$. The dynamical laws then guarantee
that~\eqref{equi} holds for all times $t$, a feature that is denoted
as \textit{equivariance}. So, Born's rule is replaced by relation~\eqref{equi}
which, in lack of a derivation, has the status of a \emph{hypothesis},
and it is called the \emph{quantum equilibrium hypothesis}. From there,
with the help of the dynamical laws, the Born rule can be derived,
so it no longer exists as an additional postulate. There are attempts
to derive the quantum equilibrium hypothesis at least in an approximative
manner. Valentini has shown that any arbitrary initial probability
density on the configuration space becomes eventually indistinguishable
from $|\Psi_{t}|^{2}$ at a coarse-grained scale \citep{Valentini1991a}.
His theorem is partly analogous to Boltzmann's famous H-theorem, which
motivates Valentini to name his theorem the \emph{subquantum H-theorem}.
Dürr, Goldstein and Zanghi propose to consider the quantum equilibrium
as a feature of \emph{typical} initial configurations \citep{Durr_et_al_1992}.
However, I do not consider these justifications of the quantum equilibrium
hypothesis satisfactory, for reasons that go beyond the scope of this
paper and have to be outlined separately. In the world continuum theory,
there is no quantum equilibrium hypothesis, and all probabilities
emerge as epistemic probabilities caused by the ignorance of the observer
about which world it is that they live in.

The second issue with Bohmian mechanics may at first sight appear
rather harmless, but on a closer look it develops considerable destructive
power: the issue of \emph{empty branches}. These are the components
of the post-measurement state that do not guide any particles because
they do not have the \textit{actual configuration} in their support.
At first sight, the empty branches do not appear problematic but on
the contrary very \textit{\emph{helpful}} as they enable the theory
to explain \textit{\emph{unique outcomes}} of measurements. Also,
they seem to explain why there is an effective ``collapse of the
wavefunction'', like in standard quantum mechanics. On a closer view,
though, one must admit that these empty branches \emph{do not actually
disappear}. As in Bohmian mechanics the wavefunction is taken to describe
a \emph{really existing field}, all their branches \emph{really exist}
and will evolve forever by the Schrödinger dynamics, no matter how
many of them will become empty in the course of the evolution. This
circumstance has led David Deutsch to famously phrase that ``pilot-wave
theories are parallel-universes theories in a state of chronic denial''
(\citealp{Deutsch1996}; this is a comment on \citealp{Lockwood1996};
for a follow-up discussion see \citealp{Valentini2008,Brown2009}).
Every branch of the global wavefunction describes a complete world
which is, according to Bohm's ontology, only a \emph{potential world}
that \textit{\emph{would}} be actual if only it were filled with particles.
Exactly one branch at a time is occupied by particles, thereby representing
the \emph{actual world}, while all other branches, though really existing
as part of a really existing wavefunction, are empty and thus describe
some sort of ``zombie worlds'' with potential planets, oceans, trees,
cities, cars, and potential people who would talk like us and behave
like us, but who do not \emph{actually exist}. The empty branches
of the wavefunction are still \textit{real}, because the entire wavefunction
is considered to be \textit{real, }but they have no further influence
on the particles. So, is there any convincing justification to consider
empty branches still as \textit{real}, beyond mere stipulation? Why
is the \textit{effective} collapse of the wavefunction not a \textit{real}
collapse? If a many-worlds theory may be accused of ontological extravagance,
then Bohmian mechanics may be accused of ontological \emph{wastefulness}.
Because, \textit{\emph{on top of}}\textit{ }the ontology of the wavefunction
with all its branches comes the additional ontology of particles,
whose actual configuration degrades the reality of most of the branches
of the wavefunction into mere potentiality. Yet, the actual configuration
is \textit{\emph{never needed}}\textit{ }for the calculation of the
statistical predictions in experimental reality, for these can be
obtained by mere wavefunction algebra. In the world continuum theory,
in contrast, there is no such thing as \textit{the actual configuration}.
All configurations in the support of the wavefunction are equally
real, and the objective description of the universe does not need
a specification of one of these configurations being the actual one.
Probability comes into play not as the ignorance of observers about
which configuration is \textit{the actual one}, but rather as their
ignorance about which configuration is \textit{the configuration of
their world}, in a sense that is precisely specified by the here-provided
logical framework.

The third issue with Bohmian mechanics is the separate existence of
wavefunction and particles, and the strange way that these entities
interact with each other. While the wavefunction acts upon the particles,
the particles do not act upon the wavefunction. So actually, there
is no \textit{inter}action between the wavefunction and the particles;
the relation is asymmetric. However, although the particles never
act back on the wavefunction, it is always the particles that define
the unique outcome of measurements; it is the particles that define
which branch of the wavefunction is the relevant one while the other
branches become empty and can be neglected. So, although the particles
have the last word, they are yet so powerless that they cannot even
act upon the wavefunction. In the world continuum theory, in contrast,
there is no separate existence of wavefunction and particles, and
no bizarre one-way ``inter''-action between these entities. There
is only one unified physically existing entity, the world continuum,
and the wavefunction is just an abstract mathematical construct that
contains all information needed to determine the \textit{form} of
the world continuum. Moreover, in the world continuum theory, particles
are not point-like entities like in Bohmian mechanics, but rather
all particles together constitute a continuous substance, the world
continuum, and it is just the time-world instances of the world continuum
that appear as discrete point-like particles. In the world continuum
theory, thus, the dualism between the continuum and the discrete,
between wave and particle, is resolved in a unique fashion. Concluding,
although being formally equal to Bohmian mechanics in many aspects,
the world continuum theory draws an entirely different picture of
the physically existing universe.

\subsection{Mind and world\label{sec:Mind-and-world}}

I will not say anything about the relation between mind and world
that goes beyond the absolute minimum. ``If there exist many worlds'',
the opponent might ask, ``then why is it that we experience only
one of them?'' The simplest answer is: ``For reasons analog to those
that make us experience only one time point as the present time''.
A more elaborate answer requires one to agree that perception is a
conscious process, and that conscious processes \textit{supervene}
on physical processes. That is to say, a conscious process is taken
to be \textit{uniquely determined} by physical processes. If all physical
processes are given then all conscious processes are also given. As
all physical processes are determined by the movement of the particles
in the universe, a conscious process is eventually determined by the
temporal evolution of the configuration of the universe, which is
given in each world by the universal trajectory in that world. In
some world $w_{1}$ the universal trajectory is $\gamma_{1}$, in
some other world $w_{2}$ the universal trajectory is $\gamma_{2}$.
Hence, in some world $w_{1}$ at time $t$, Joe is in a certain mental
state $M_{1}$ uniquely determined by the trajectory point $\gamma_{1,t}$,
and in some other world $w_{2}$ at the same time $t$, Joe is in
some other mental state $M_{2}$ uniquely determined by the trajectory
point $\gamma_{2,t}$. In no world, however, Joe is in the mental
states $M_{1}$ and $M_{2}$ together at the same time. In the same
manner that traffic lights cannot be green and red at the same time
in the same world, Joe cannot be in the mental states $M_{1}$ and
$M_{2}$ at the same time in the same world. Roughly speaking, Joe
cannot be experiencing more than one world, because his experience
is part of the world, so in different worlds Joe has different experiences.
Say, at time $t$ Joe is in one world $w_{1}$ experiencing green
traffic lights, and in another world $w_{2}$ he is experiencing red
traffic lights. Although there is no logical contradiction in Joe
experiencing both green and red traffic lights at the same time, because
he does so in different worlds, it may seem less confusing to talk
about different \textit{world-instances} of Joe experiencing the traffic
lights either as red or as green at time $t$. So rather than speaking
of Joe being in a mental state $M_{1}$ at time $t$ and in world
$w_{1}$, while being in another mental state $M_{2}$ at the same
time $t$ in another world $w_{2}$, one may equivalently speak of
one world-instance of Joe, namely $\text{Joe}{}_{1}\equiv\text{Joe}@(w_{1})$
being in one mental state $M_{1}$ at time $t$, and another world-instance
of Joe, namely $\text{Joe}{}_{2}\equiv\text{Joe}@(w_{2})$ being in
another mental state $M_{2}$ at time $t$. There is no instance of
Joe that is in both mental states $M_{1}$ and $M_{2}$ at time $t$.

It is difficult to decide whether and to what extent there might be
causal connections between worlds, like there are causal connections
from past world-instances to future world-instances. If there were
such inter-world causation, there would be the possibility of imprints
in an observer's brain in a specific world caused not only by events
occurring in past instances of that world, but also by events in parallel
worlds. This would entail the possibility of having a ``counterfactuals
sense'' analogue to having a memory. The many-interacting-worlds
(MIW) approach by \citet{Hall_et_al_2014}, which bears many similarities
to my approach, seems to support the possibility of inter-world causation.
However, as I will substantiate further on, I would rather not regard
the relation between worlds as \textit{interaction} but as \textit{interference},
which may have consequences on the possibility of having a counterfactuals
sense. In any case a closer analysis, which also entails taking into
account different concepts of causation, is needed though clearly
outside the scope of this paper.

\subsection{Tipler's approach\label{sub:Tipler's-approach}}

There is a very interesting formulation of quantum mechanics by \citet{Tipler2006}
which seems to be similar in spirit to the ideas proposed here. The
author explicitly writes (\emph{ibid}, page 1): 
\begin{quote}
The key idea of this paper is that the square of the wave function
measures, not a probability density, but a density of universes in
the multiverse.
\end{quote}
Unfortunately, though, Tipler deviates from his initial conception
when, for example, he later writes (\emph{ibid}, page 4): 
\begin{quote}
In the case of spin up and spin down, there are only two possible
universes, and so the general rule for densities requires us to have
the squares of the coefficients of the two spin states be the total
number of effectively distinguishable – in this case obviously distinguishable
— states. 
\end{quote}
Such statement is hard to understand. If the number of universes (or
``worlds'', as the author also calls them elsewhere in the paper)
is two, then what does it mean to ``have the squares of the coefficients
of the two spin states be the total number of effectively distinguishable
{[}...{]} states''? The word ``state'' seems to refer to a universe,
or world, and within the same sentence also to something else. How
many ``states'' or ``universes'' are there, in that situation,
two or infinitely many? Such ambiguity and vagueness about the ontological
meaning of the terms ``states'', ``branches'', ``worlds'', ``universes'',
is somewhat idiosyncratic for Everett-type theories. In the here-proposed
theory, in contrast, worlds correspond to well-defined trajectories
in configuration space and hence their total number is always uncountably
infinite, and spin states are just components of the wavefunction
and not labels for, or representatives of, worlds. It seems that after
all Tipler sticks to the Everettian ontology of \emph{branches} rather
than to a continuous multiplicity of worlds, in contrast to what the
author's initial statement seems to suggest. Other strong indicators
for that conclusion are that in Tipler's analysis the universes still
split, or ``differentiate'', as the author also calls it, and that
he explicitly writes ``the sums in (15) {[}...{]} are in 1 to 1 correspondence
with real universes'', where the referenced formula involves a decomposition
of the wavefunction into spin states. Another fundamental difference
to the world continuum theory concerns the justification of probabilities.
Tipler writes (\emph{ibid}, page 1--2): 
\begin{quote}
The probabilities arise because of the existence of the analogues
of the experimenters in the multiverse, or more precisely, because
before the measurements are carried out, the analogues are `indistinguishable'
in the quantum mechanical sense. Indistinguishability of the analogues
of a single human observer means that the standard group transformation
argument used in Bayesian theory to assign probabilities can be applied.
I show that the group transformation argument yields probabilities
in the Bayesian sense, and that in the limit of an infinite number
of measurements, the relative frequencies must approach these probabilities.
\end{quote}
Different from this rather sophisticated justification, the probabilities
in the here-proposed theory derive from a straightforward generalization
of the Laplacian rule to a continuum of possibilities. I conclude
that Tipler's approach is conceptually different from mine.

\subsection{The MIW approach\label{sub:many-interacting-worlds}}

Bohmian mechanics can be formulated, so that the Bohm equation \eqref{eq:guidingeq}
obtains a Newtonian form
\begin{equation}
m_{n}\ddot{\boldsymbol{\gamma}}_{n}=-\boldsymbol{\nabla}_{n}(V_{t}+Q_{t}),\label{eq:Newtoneq}
\end{equation}
for $n=1,\ldots,N$, involving the so-called \textit{quantum potential}
\begin{equation}
Q_{t}=\sum_{n=1}^{N}-\frac{\hbar}{2m_{n}}\frac{\boldsymbol{\nabla}_{n}^{2}\sqrt{\rho_{t}}}{\sqrt{\rho_{t}}},\label{eq:quantum-potential}
\end{equation}
and with the initial velocity of the particles being restricted by
\begin{equation}
\dot{\boldsymbol{\gamma}}_{n,0}=\frac{\boldsymbol{j}_{n,0}(\gamma_{0})}{\rho_{0}(\gamma_{0})}.\label{eq:Newtoninitial}
\end{equation}
In this formulation, which was presented by Bohm himself \citeyearpar{Bohm1952,Bohm1952a},
it becomes explicit how the motion of the particles is affected not
only by classical forces but also by non-classical forces generated
by the quantum potential, which vanishes in the classical limit $\hbar\rightarrow0$.
Since the quantum potential is a function of the density $\rho_{t}=|\Psi_{t}|^{2}$,
Bohm interpreted the non-classical force as the force that the wavefunction
exerts on the particles. 

In a very interesting approach, \citet{Hall_et_al_2014} propose to
replace the density $\rho_{t}$ in \eqref{eq:quantum-potential} with
the empirical density $\rho_{t}^{K}$ of a discrete number of trajectories,
identified as \textit{worlds}, so that
\begin{equation}
\rho_{t}^{K}(q)=\frac{1}{K}\sum_{k=1}^{K}\delta(q-\gamma_{k,t}),
\end{equation}
where $K$ is the number of worlds. That way, the one-way interaction
between the wavefunction and the particles becomes a two-way interaction
between \textit{worlds}, in that the evolution of each world configuration
$\gamma_{k,t}$ is directly determined by all other world configurations.
If the worlds are distributed by a probability density equal to $|\Psi_{t}|^{2}$
then for $K\rightarrow\infty$ the world density $\rho_{t}^{K}$ approaches
(in a distributional sense) the standard density, $\rho_{t}^{K}\rightarrow|\Psi_{t}|^{2}$,
hence the theory would reproduce the predictions of standard quantum
mechanics in the limit of infinitely many worlds. 

I find this proposal very fascinating, as it shows that quantum effects
can be explained by the interaction between \textit{physically existing
worlds}. I have only three criticisms. One is that I would prefer
to call the relation between worlds not \textit{interaction} but rather
\textit{interference}, because the difference between interaction
and interference seems to me a categorical one. Interaction is a relation
between \textit{systems}, while interference is a relation between
\textit{worlds}, that is, between trajectories of one and the same
system. For example, an electron in a double-slit experiment \textit{interacts}
with the walls of the double-slit and with the detection screen behind
the slits. On the other hand, a trajectory of the electron passing
through one slit \textit{interferes} with the trajectories of the
same electron passing through the other slit. If one of the slits
is closed, then the trajectories of the electron cannot pass the region
covered by the closed slit, and all trajectories go through the open
slit, interfering with each other and producing a typical single-split
diffraction pattern on the screen when the experiment is repeated
many times. When the second slit is opened, the trajectories of the
particle may pass both slits, and behind the double-slit they \textit{interfere}
with each other, producing a typical double-slit interference pattern
on the screen. In each world there is only one trajectory passing
through only one of the slits, but since the course of that trajectory
is influenced by other trajectories, each belonging to a separate
world, one is led to conclude that the typical wave phenomena of particles
can be interpreted as the influence that individual worlds exert on
one another. This special kind of mutual influence between worlds
is different in nature from ordinary interactions between systems
and, therefore, so my suggestion, should be distinctively identified
as \textit{interference}. 

My second criticism of the approach of \citeauthor{Hall_et_al_2014}
is, of course, that they assume a finite number of worlds. For one
part, if the predictions of standard quantum mechanics are reproduced
in the limit of infinitely many worlds, then why not taking infinitely
many worlds? The here-proposed theory already involves right from
the start an infinite number of worlds while being mathematically
much simpler than the proposal by \citeauthor{Hall_et_al_2014}. The
world continuum theory may be regarded as the limiting theory obtained
from the many-interacting-worlds (MIW) theory of \citeauthor{Hall_et_al_2014}
by taking the number of worlds to infinity. 

My third criticism is that the MIW relies, like Bohmian mechanics,
on a probability assumption. Only if it is assumed that the finitely
many worlds are at $t=0$ spatially distributed by $|\Psi_{0}|^{2}$,
in the sense of a \textit{probability} distribution, the predictions
of standard quantum mechanics are approximately reproduced. Thus,
it is impossible to conceptually \textit{derive} probability from
the MIW. Nonetheless, the authors claim that the worlds are equally
weighted, so that a probability argument analog to that of Laplace
would be applicable (\textit{ibid}, p 041013):
\begin{quote}
It is a remarkable feature of the MIW approach that only a simple
equal weighting of worlds, reflecting ignorance of which world an
observer occupies, appears to be sufficient to reconstruct quantum
statistics in a suitable limit.
\end{quote}
At a closer look, there are two types of probability brought into
play here. One is the probability $p_{k}$ that world $k$ is the
world of the observer, and the other is the probability density $\rho_{k,0}(q)$
to find at time $t=0$ the configuration of world $k$ at $q$. Due
to the observer's complete ignorance which world it is that they live
in, the probability $p_{k}$ is equally weighted, so $p_{k}=1/K$.
In the same vein, the probability density $\rho_{k,0}$ is assumed
to be equal for all worlds, so $\rho_{k,0}=|\Psi_{0}|^{2}$. Importantly,
while each world is given the same probabilistic weight with respect
to its bare identity, which can be justified by an ignorance assumption,
its possible configurations are \textit{not} given the same probabilistic
weight. Some configurations are more probable than others, and there
is no simple ignorance argument for such a non-trivial probability
assumption. So my criticism is that the authors conceal a non-trivial
and not justified probability assumption concerning the configuration
of the world of the observer by a trivial and justified probability
assumption concerning the bare identity of the world of the observer.
They do not address the following question: why are some configurations
more probable than others, and where does that non-uniform probability
come from? As a side remark: it does not help in this respect to formally
parameterize the worlds by some parameter $C$, so that $C$ becomes
uniformly distributed, as is done by Schiff and Porier in their approach,
which is conceptually very close to mine \citep{Schiff_et_al_2012}.
For, the explanatory burden is now shifted towards answering the question:
Where does that parameterization come from? It appears \textit{ad
hoc} to introduce a specific parameterization of worlds only to get
the probability distribution over the new parameter $C$ uniform,
as long as there is no \textit{physical} justification for doing so.
Why should the observer apply the Laplacian rule of indifference towards
some fictitious parameter $C$, when it is really and only the positions
of particles, that is, the \textit{configuration} of the world, which
he can (in principle) measure? The parameter $C$ is a merely theoretic,
not an empirical, quantity that depends on the probability distribution
over the possible world configurations, so it would be circular to
use it for a derivation of that very probability distribution. 

As I have argued in section \ref{sub:Beyond-Laplace}, there is an
epistemological explanation for the occurrence and the numerical values
of quantum mechanical probabilities when interpreting $|\Psi_{t}|^{2}$
not as a probability density but rather as the \textit{substantial
density} of a physically existing continuum of worlds. Probability
can then be conceptually derived, in a manner analogous to the classical
derivation by Laplace, from both the physical facts (the empirically
given distribution of the world continuum) and the observer's indifference
towards which one out of the continuously many worlds is actually
the world of the observer. The configuration of the world of the observer
is more likely to be found near the location $q$ than near another
location $q'$ exactly if $\rho(q)>\rho(q')$, since at $q'$ the
worlds are \textit{more densely packed} than at $q'$. As a discrete
analogue: It is more probable to find a particular needle out of many
needles in a haystack if you look at those places where there are
more needles. You do not even have to \textit{know} the distribution
of needles in the stack. The probability to find your needle \textit{is},
as a matter of physical \textit{fact}, bigger when you (accidentally)
look at places where there are more needles. The probability itself
is epistemic, but it derives from ontological facts. Now, the interpretation
of $|\Psi_{t}|^{2}$ as a substantial density of worlds is impossible
when considering an only finite number of worlds. As such, the substantial
density is an empirical distribution, and if there were only finitely
many worlds, the empirical distribution had to be a finite sum of
delta functions, which does not apply for $|\Psi_{t}|^{2}$.

\subsection{Newtonian QM}

\citet{Sebens2014} considers a large but finite number of worlds
in his proposal of a \textit{Newtonian QM}. He completely removes
the wavefunction from the theory, leaving only the density $\rho=\rho_{t}(q)$
and the velocity field $v=v_{t}(q)$, which together yield the flow
$j=\rho v$. With the wavefunction having been removed, the Newtonian
equation \eqref{eq:Newtoneq} for the movement of the particles cannot,
as in Bohmian mechanics, be interpreted as the action of a classical
potential and a quantum field on each particle. Instead, the equation
seems to describe the action of a classical potential and a classical
continuous fluid described by the density $\rho$ on each particle.
However, according to Sebens' construction of the theory, there is
no continuous fluid but just a finite number of worlds, and the density
$\rho$ is just an approximation obtained by averaging over a large
enough number of worlds. Hence, as Sebens points out, \eqref{eq:Newtoneq}
cannot be considered a fundamental equation, so that empirically testable
deviations may occur. Furthermore, as the wavefunction is no longer
there to put restrictions on the density and the velocity field, there
are combinations of the latter two that do not correspond to any underlying
wavefunction and, therefore, do not represent empirically adequate
scenarios. To restore the empirical adequacy of the theory, Sebens
introduces an additional postulate, the ``quantization condition'',
which holds that
\begin{equation}
\oint\left\{ \sum_{k}\left[m_{k}\boldsymbol{v}_{k}\cdot d\boldsymbol{l}_{k}\right]\right\} =nh.
\end{equation}
Sebens admits that he has no satisfying justification for the additional
postulate other than that it would naturally follow if one presumes
the existence of an underlying wavefunction. However, with the wavefunction
existing\textit{ in addition} to a large number of worlds, the theory
blows up to an ontologically extravagant theory, \textit{Prodigal
QM}, which does not offer interpretational advantages against standard
Bohmian mechanics. Therefore, he introduces the wavefunction again,
though holding that it does not represent a physically existing entity
itself, but rather ``a convenient way of summarizing information
about the positions and velocities of particles in the various worlds''
(\textit{ibid}, p. 9).

I find Sebens' approach and his interpretational analysis particularly
useful in that it clearly shows at which points problems occur that
have to be addressed by additional conceptual work. My own conclusions
are: 1) to accept that there is a continuous infinity of worlds corresponding
to Bohmian trajectories (here I differ from Sebens), and 2) to interpret
the wavefunction not as a physically existing field in addition to
particles and worlds, but rather as a generating function for, and,
therefore, as a mathematical representation of, a physically existing
entity that already \textit{entails} particles and worlds (which is
more or less what Sebens also concludes).

\subsection{Poirier and Schiff's approach}

Perhaps the approach closest to my proposal is Poirier and Schiff's
reformulation of Bohmian mechanics \citep{Poirier2010,Schiff_et_al_2012}.
In a sense, their formulation and mine are two formulations of the
same theory, only emphasizing different aspects of it. While my formulation
puts a focus on the interpretation of the density function $\rho$,
as well as on the ontological and epistemological aspects of the theory,
Poirier and Schiff's formulation explores formal aspects and their
physical relevance. Their main result is that one may completely get
rid of the wavefunction, leaving only the trajectory function $\xi=\xi_{t}(q)$
and the initial density function $\rho_{0}=\rho_{0}(q)$ as a complete
representation of the history of a closed system\footnote{Here and in the following I adapt their notation to mine for the sake
of easier comparison.}. At first, \citet{Poirier2010} showed that the entire dynamics of
a 1D system is captured by a partial differential equation (PDE) for
the trajectory function $\xi=\xi_{t}(q)$, involving temporal derivatives
$\dot{\xi},\ddot{\xi}$ as well as spatial derivatives $\xi',\xi'',\xi'''$.
He then showed that the system dynamics can be expressed in terms
of a minimal-action principle on a generalized Lagrangian density
of the form
\begin{equation}
\mathcal{L}[q,\xi,\dot{\xi},\xi',\xi'',\xi''']=\rho_{0}(q)L[q,\xi,\dot{\xi},\xi',\xi'',\xi'''],
\end{equation}
with the Lagrangian
\begin{equation}
L[q,\xi,\dot{\xi},\xi',\xi'',\xi''']=\frac{1}{2}m\dot{\xi}^{2}-V[\xi]-Q[q,\xi',\xi'',\xi'''],
\end{equation}
and a rather complicated quantum potential $Q$. \citet{Schiff_et_al_2012}
generalized the theory to the many-$D$ case, and they also considerably
simplified $Q$ and the PDE by re-parametrizing the trajectory function
$\xi_{t}(q)$ to $\xi_{t}(C)$ with the parameter $C$ being chosen
so that it uniformizes the probability density $\rho_{0}(q)$ to $\rho_{C}(C)\equiv1$,
where $\rho_{C}(C)\,dC=\rho_{0}(q)\,dq$. The relevance of their work
to the understanding of quantum mechanics is profound. For one part
it shows that a complex-valued wavefunction need not be considered
an indispensable ingredient to quantum theory. For another part, and
alongside with \citet{Hall_et_al_2014} and \citet{Sebens2014}, the
work of Poirier and Schiff shows that quantum phenomena can be regarded
as being caused by the interaction (or, as I would prefer to call
it, by the interference) of a multitude of equally existing trajectories.
For, if one were to reject the idea of equally existing trajectories,
then one would have to justify how and in what sense a lonely existing
trajectory can possibly interact (or interfere) with other, non-existing,
trajectories. 

Although in my approach the wavefunction is still there, it is degraded
to a mere ``generating function'' for the physically existing trajectories,
with no reality on its own. In this respect, both of our approaches
go beyond what Bohm and his adherents had, respectively, have in mind.
The trajectories are the same in all three theories, and they form
a continuum, so this is common ground, too. 

However, what Porier and Schiff did not fully work out, and what is
more closely addressed in my approach, is the physical meaning of
the collection of trajectories and of the density function. Porier
and Schiff refer to the collection of trajectories as an \textit{ensemble},
which, however, remains a rather ambiguous reference. In statistical
theories, an \textit{ensemble} is a hypothetical, infinite collection
of independent realizations of a random variable, and the probability
distribution associated with the outcomes of the variable governs
the relative frequencies of individual realizations within the ensemble.
A statistical ensemble is in any case countably infinite, even for
a finite number of possible values of the random variable under consideration.
For example, the ensemble associated with a fair coin would be an
infinite sequence of heads and tails, with the relative frequency
of heads within a partial sequence converging towards one half as
the length of the partial sequence goes to infinity. This is certainly
not the concept of \textit{ensemble} that Poirier and Schiff intend
to invoke. Rather, their ``ensemble'' of trajectories is the uncountably
infinite collection of \textit{all possible} trajectories, which are
parameterized solutions of one and the same differential equation.
In terms of a statistical theory, their ``ensemble'' would rather
represent the (uncountable) collection of all possible \textit{outcomes},
not a (countable) collection of \textit{realizations}, of a random
trajectory. The density function $\rho$ would then represent the
probability density on the collection of possible trajectories, with
only one trajectory being realized. Poirier seems to sympathize with
this view, which coincides with the standard view on Bohmian mechanics,
when he writes \citep[p. 14]{Poirier2010}:
\begin{quote}
Perhaps the most natural association between the present formulation
and existing interpretations of quantum mechanics would be with the
statistical or ensemble interpretation {[}\textit{cit.}{]}, whose
most famous proponent was Einstein himself. Indeed, the trajectory
density weighting function, $\rho_{0}(x_{0})$, may be regarded as
a classical statistical probability distribution {[}...{]}.
\end{quote}
At several places in the paper (\textit{ibid}, pp. 4, 9, 10, 13, 14),
Poirier calls $\rho$ a ``trajectory density weighting'', but he
does not explicate what this actually \textit{means}. At other places
again (\textit{ibid}, pp. 6, 8, 9, 11, 13, 14), the author stays with
the standard notion of ``probability''. Also, there is no clear
commitment as to how his approach relates to the existence of parallel
\textit{worlds}. On page 14 we read:
\begin{quote}
If one presumes objective existence for a single trajectory only,
then the remaining trajectories in the ensemble must be regarded as
``virtual'', in some sense. On the other hand, one might prefer
to regard all trajectories in the quantum ensemble as equally valid
and real. It is hard to imagine how this could be achieved, without
positing that each trajectory inhabits a separate world. It must be
emphasized, however, that this version of the many worlds interpretation
would be very different from the standard form {[}\textit{cit.}{]}.
\end{quote}
In their later paper Schiff and Poirier still leave open how to interpret
their formal framework \citep[p. 031102-4]{Schiff_et_al_2012}:
\begin{quote}
Regarding interpretation, we draw no definitive conclusions here.
However, it is clearly of great significance that the form of $Q$
can be expressed in terms of $x$ and its $C$ derivatives—implying
the key idea that the interaction of nearby trajectories, rather than
particles, is the source of all empirically observed quantum phenomena
(suggesting a kind of “many worlds” theory, albeit one very different
from Ref. 5).
\end{quote}
Somewhere else again in the paper, the authors explicitly speak of
$\rho_{0}(x_{0})$ as of a ``probability density'' (\textit{ibid},
p. 031102-2). Even if one is willing to read their approach in the
spirit of a many-worlds theory, in such a way that all members of
the collection of trajectories (their ``ensemble'' of trajectories)
are taken to equally \textit{exist}, it remains unclear how to interpret
the density function $\rho$. If it were not a probability density,
what kind of density could it possibly represent? Calling it a \textit{density
weighting} at times \citep[pp. 5, 9, 10, 13, 14]{Poirier2010} does
not really help as it still leaves open what kind of ``weighting''
is mediated by $\rho$ in their theory (if not a \textit{probability}
weighting).

I conclude that the approach by Poirier and Schiff bears great similarities
to my proposal, but its physical interpretation remains too vague
to be considered a complete equivalent. My personal view is that Poirier
and Schiff's approach and mine \textit{complement} each other in that
those questions having been left open by either approach are addressed
by the other and that they are essentially about the same \textit{theory}.

\subsection{Why should the worlds form a continuum?\label{sec:Can-the-worlds}}

In a recent paper, \citet{Vaidman2014} discusses numerous formulations
and interpretations of quantum mechanics, and in particular the idea
put forward in \citet{Tipler2006} and in \citet{Bostrom2012}, of
a continuum of worlds existing in parallel. Although Vaidman strongly
argues for a multiplicity of worlds, he rejects the idea that there
is a continuum of them. Rather, so he argues, there can only be a
large but finite number of parallel worlds. He writes \citep[p. 24]{Vaidman2014}:
\begin{quote}
If in any region of the configuration space there is an infinity of
worlds, we cannot say that it is smaller or larger than in another
region. {[}…{]} Boström apparently adopts the concept of the measure
of existence. His proposal for “volume of worlds” (16) includes also
the “weight” of each world $|\Psi_{t}(q)|^{2}$. But then, I do not
see how it can be considered as a density of worlds. As far as I can
understand, for an infinite number of worlds there is no mathematical
formalism which can provide a “density of worlds” picture. 
\end{quote}
In the same context, with regard to Albert and Loewer's \textit{many-minds}
approach \citep{Albert_et_al_1988}, where the authors propose a continuum
of minds together with a measure on the totality of minds, Vaidman
writes: ``There is no mathematical formalism which provides such
a measure''. In line with Vaidman, \citet{Hall_et_al_2014} and \citet{Sebens2014}
propose a large but finite set of worlds, which they identify with
Bohmian trajectories. Although they admit that only in the limit of
a continuum of worlds the predictions of standard quantum mechanics
are exactly reproduced, they insist on a finite set. \citet[p. 14]{Sebens2014}
writes:
\begin{quote}
The meaning of $\rho$ becomes unclear if we move to a continuous
infinity of worlds since we can no longer understand $\rho$ as yielding
the proportion of all worlds in a given volume of configuration space
upon integration over that volume. There would be infinite worlds
in any finite volume (where $\rho\neq0$) and infinite total worlds.
If $\rho$ doesn’t give the proportion of worlds in a region, it is
unclear why epistemic agents should apportion credences as recommended
in the previous section. So, the continuous variant, if sense can
be made of it, faces the quantitative probability problem.
\end{quote}
I do not share these worries. The mathematical formalism that I and
certainly also the other cited authors tacitly refer to when considering
a continuum of worlds (or minds) endowed with a measure, is \textit{measure
theory}. Briefly, a \textit{measure} is a function $\mu$ on a collection
$\Sigma$ of subsets of a set $\mathcal{X}$, which maps each subset
$X\in\Sigma$ to a non-negative number $\mu(X)$ called the \textit{measure}
of $X$. The collection $\Sigma$ is called a \textit{$\sigma$-algebra}
of the set $\mathcal{X}$, and its elements are called \textit{$\mu$-measurable
subsets} of $\mathcal{X}$. Both the $\sigma$-algebra and the measure
fulfill certain requirements. The cardinality of $\mathcal{X}$ does
not matter; it might be finite, countable, continuous, or of even
higher cardinality. Measure theory forms the basis of integral theory,
for the Lebesgue integral $\lambda(X)=\int_{X}dx$ can be regarded
as a special measure on the vector space $\mathcal{X}=\mathbb{R}^{N}$.
Any measure on a Lebesgue-measurable space can be associated with
a \textit{density} $\rho$, such that $\mu(X)=\int_{X}dx\,\rho(x)$,
with the Lebesgue measure itself having the trivial density $\rho\equiv1$.
As has already been discussed at some length, my proposed interpretation
of a regular, i.e non-singular, density in physical space is that
of a distribution of a continuous substance (a fluid, say) across
the space $\mathcal{X}$, so that $\mu(X)$ yields the \textit{amount
}of substance contained within a given region $X\subset\mathcal{X}$.
In regions where the integrated density is higher, there is \textit{more
substance} than in other regions where the integrated density is lower.
According to my proposal, the collection of all trajectories that
solve the Bohm equation \eqref{eq:guidingeq} constitutes a continuous
substance distributed in configuration space, with each trajectory
uniquely corresponding to an individual \textit{world}. So $\mu_{t}(X)$
measures the amount of worlds whose configuration is contained at
time $t$ within the region $X$.

If the density within a certain region $X$ is smaller than unity,
one might be tempted to assume that in $X$ there must be \textit{holes},
that is, places where there are no worlds. However, this is not necessarily
the case. The worlds form a \textit{continuum}, and a continuum does
not have to have holes for its density to be decreased. As a simple
toy model consider a set $\mathcal{X}=[0,1]$ of ``worlds'' being
embedded into a set $\mathcal{Y}=[0,1]$ of ``configurations'' by
the function $f(x)=x^{2}$. The function $f$ is a bijection here,
so the set $\mathcal{Y}$ is entirely filled with points from $\mathcal{X}$
without any holes. Set theoretically, there are exactly as many points
in $\mathcal{X}$ as in $\mathcal{Y}$, namely $\mathfrak{c}=2^{\aleph_{0}}$.
Still, the density of points taken from $\mathcal{X}$ and embedded
into $\mathcal{Y}$ is not uniform; there are ``more'' points (in
the sense of a higher measure) from $\mathcal{X}$ contained in the
lower end of $\mathcal{Y}$ than contained in the upper end. If $\nu$
is a measure on $\mathcal{X}$ then the measure $\mu$ on $\mathcal{Y}$
derived from $\nu$ is given by
\begin{equation}
\mu(Y)=\nu(\{x\in\mathcal{X}\mid f(x)\in Y\}).
\end{equation}
So if we take the worlds in $\mathcal{X}$ to be uniformly distributed
according to $\nu(X)=\int_{X}dx$, then we can explicitly calculate
the measure $\mu$ on the configurations in $\mathcal{Y}$ as
\begin{align}
\mu(Y) & =\int_{[0,1]}dx\int_{Y}dy\,\delta(y-f(x))\\
 & =\int_{Y}dy\,\frac{1}{2\sqrt{y}},
\end{align}
and so the configurations of the worlds in $\mathcal{X}$ are distributed
within $\mathcal{Y}$ according to the non-uniform density $\rho_{Y}(y)=\frac{1}{2\sqrt{y}}$.
Then, the subjective probability that our actual world has a configuration
within a certain region $Y\subset\mathcal{Y}$ is given by the ratio
of the amount of worlds whose configuration lies within $Y$ divided
by the amount of all worlds, so $P(Y)=\mu(Y)/\mu(\mathcal{Y})$. The
probability that a particular world (for example \textit{our} world)
has a configuration within the interval $[0,a]$ for $0\leq a\leq1$
is given by $P([0,a])=\sqrt{a}$. The toy model illustrates how the
notion of a density of worlds whose configurations are distributed
across configuration space makes sense also in the case of a continuum
of worlds. 

Besides these rather mathematical aspects, why should one consider
the set of worlds to form a continuum in the first place? Within the
framework of my proposal, the answer is quite simple: because there
is a continuum of trajectories that solve the Bohm equation \eqref{eq:guidingeq}.
If only finitely many worlds were allowed, which of these trajectories
should be taken as real, and for what reason? Logically, there is
an infinite number of possible paths that the configuration of the
universe may take. Invoking a differential equation (the Bohm equation)
selects the physically possible paths from the logically possible
ones, and the so-selected paths derive from one and the same wavefunction
$\Psi_{t}$ representing the state of the universe. If only $\Psi_{t}$,
and nothing more, is considered to represent a complete description
of the universe, then all paths deriving from $\Psi_{t}$ must be
considered as real. Discarding all but a finite number of paths is
not only unnecessary and \textit{ad hoc}, but it requires an additional
postulate. And, if one allows for only a finite number of worlds then
why not directly postulating that there is just \textit{one} world,
as is done in Bohmian mechanics? Finitely-many-worlds theories based
on the conception of worlds as trajectories in configuration space,
are axiomatically \textit{more} demanding than the world continuum
theory, as they must specify, in addition to the wavefunction, the
trajectories $\gamma_{1},\ldots,\gamma_{K}$ of those worlds that
are taken as real, so that the objective state of the universe at
time $t$ is given by
\begin{equation}
\left(\begin{array}{c}
\Psi_{t}\\
\gamma_{1,t},\ldots,\gamma_{K,t}
\end{array}\right),
\end{equation}
for $K$ representing the number of objectively existing worlds, and
with $\gamma_{k}=(\boldsymbol{\gamma}_{k,1},\ldots,\boldsymbol{\gamma}_{k,N})$
being the trajectory of the $k$-th world, whose configuration at
time $t$ is given by $\gamma_{k,t}$. In the case $K=1$ one obtains
Bohmian mechanics. The here-proposed world continuum theory, in contrast,
just needs the wavefunction $\Psi$ to mathematically represent a
continuum of trajectories and hence worlds, which are the solutions
of the Bohm equation \eqref{eq:guidingeq}. Therefore, in the world
continuum theory, just as in Everettian mechanics, the complete history
of the universe is specified by the wavefunction alone. In contrast
to Everettian mechanics, however, there is no ambiguity concerning
the identity of the worlds, as they are given by precisely defined
trajectories. To explain the empirical fact of the unique outcomes
of measurements, the world continuum theory needs an \textit{actual
configuration}, just like Bohmian mechanics, but it is taken as \textit{epistemic}
rather than \textit{ontic}. It specifies not \textit{the} actual configuration
of the universe, but rather the configuration that is actual \textit{to
a particular instance of an observer}. With respect to \textit{his
or her} experience, and only so\textit{, }there is a unique measurement
outcome and an actual configuration of the universe. In essence, a
complete objective description of Nature that complies with the empirical
facts, requires also taking into account the subjective \textit{experience}
of Nature.

\section{Acknowledgments}

There have been many enlightening and stimulating discussions with
Rainer Plaga, Hrvoje Nikolic, Lev Vaidman, Delyan Zhelyasov, Jens
Eisert, and Jan Michel. I also thank the reviewer for his or her thorough
criticism that inspired considerable improvements.

\bibliographystyle{apalike2}

\end{document}